\documentclass[apsrev4-1,prb,superscriptaddress,longbibliography,twocolumn,10pt,aps]{revtex4-2}
\usepackage{colortbl}
\usepackage{color}
\usepackage{fancyhdr}
\usepackage{graphicx}
\usepackage[english]{babel}
\usepackage[utf8]{inputenc}
\usepackage[T1]{fontenc}
\usepackage{textpos}
\usepackage{mathtools,wasysym}

\usepackage{amsfonts,amsmath,amssymb,bm}
\usepackage[dvipsnames]{xcolor}
\usepackage{tabularx,hhline,tikz,multirow,array}
\usepackage{bm,enumitem}
\usepackage{dcolumn,stmaryrd}
\usepackage[colorlinks=true,pdfborder={0 0 0},linkcolor=red,citecolor = blue]{hyperref}
\usepackage[caption=false]{subfig} 

\newcommand{\zt}{$\mathbb{Z}_2\:$}
\newcommand{\uo}{$\mathrm{U}$(1) }
\newcommand{\eff}{\mathrm{eff}}
\newcommand{\mrm}{\mathrm}
\newcommand{\mbf}{\mathbf}

\newcommand{\Da}{\Downarrow}

\newcommand{\da}{\downarrow}
\newcommand{\ua}{\uparrow}

\DeclarePairedDelimiter\abs{\lvert}{\rvert}
\DeclarePairedDelimiter\ket{\lvert}{\rangle}
\DeclarePairedDelimiter\bra{\langle}{\rvert}
\DeclarePairedDelimiter\braket{\langle}{\rangle}

	\newcommand{\hexa}{\mbox{%
	\begin{tikzpicture}[baseline=-0.5ex,scale=0.4]
		\centering
		\coordinate (center) at (0,0);
		\def\radius{0.5cm}
		\def\lw{1pt}
		\def\lwa{0.1pt}
		
		\foreach \i in {0,...,5} {
			\coordinate (v\i) at ({60*\i+30}:\radius);
			\fill (v\i) circle (3pt);  
		}
		\draw[line width=\lw] (v0) -- (v1);
		\draw[line width=\lw](v2) -- (v3);
		\draw[line width=\lw](v4) -- (v5);
	\end{tikzpicture}
	}}
	\newcommand{\hexb}{\mbox{%
			\begin{tikzpicture}[baseline=-0.5ex,scale=0.4]
				\centering
				\coordinate (center) at (0,0);
				\def\radius{0.5cm}
				\def\lw{1pt}
				
				\foreach \i in {0,...,5} {
					\coordinate (v\i) at ({60*\i+30}:\radius);
					\fill (v\i) circle (3pt);  
				}
				\draw[line width=\lw](v1) -- (v2);
				\draw[line width=\lw](v3) -- (v4);
				\draw[line width=\lw](v5) -- (v0);
			\end{tikzpicture}
	}}
\allowdisplaybreaks

\begin{document}

\title{Coupling quantum spin ice to matter on the centered pyrochlore lattice}
\author{Rajah P. Nutakki}
\affiliation{CPHT, CNRS, École Polytechnique, Institut Polytechnique de Paris, 91120 Palaiseau, France.}
\affiliation{Coll\`ege de France, Universit\'e PSL, 11 place Marcelin Berthelot, 75005 Paris, France}
\affiliation{Inria Paris-Saclay, 91120 Palaiseau, France}
\affiliation{LIX, CNRS, Ecole Polytechnique, Institut Polytechnique de Paris, 91120 Palaiseau, France}
\affiliation{Arnold Sommerfeld Center for Theoretical Physics, Ludwig-Maximilians-Universit\"at M\"unchen, Theresienstr. 37, 80333 M\"unchen, Germany}
\affiliation{Munich Center for Quantum Science and Technology (MCQST), Schellingstr. 4, 80799 M\"unchen, Germany}

\author{Sylvain Capponi}
\affiliation{Univ Toulouse, CNRS, Laboratoire de Physique Th\'eorique, Toulouse,  France.}

\author{Ludovic D. C. Jaubert}
\affiliation{CNRS, Universit\'e de Bordeaux, LOMA, UMR 5798, 33400 Talence, France}

\author{Lode Pollet}
\affiliation{Arnold Sommerfeld Center for Theoretical Physics, Ludwig-Maximilians-Universit\"at M\"unchen, Theresienstr. 37, 80333 M\"unchen, Germany}
\affiliation{Munich Center for Quantum Science and Technology (MCQST), Schellingstr. 4, 80799 M\"unchen, Germany}

\begin{abstract}
The low-energy physics of quantum spin ice is known to support an emergent form of quantum electrodynamics (QED), where magnetic monopoles exist and the fine structure constant is material dependent.
In this article, we show how this QED is modified via a coupling to dynamical matter on the centered pyrochlore lattice, a structure which has recently been synthesized using metal-organic frameworks.
Specifically, we study the low-energy properties of the $S = 1/2$ quantum XXZ model on the centered pyrochlore lattice, with a focus on the sign-problem free region.
At fourth order in degenerate perturbation theory this model hosts
a quantum spin liquid distinct from the well-known U(1) quantum spin ice on the pyrochlore due to the presence of dynamical matter in the ground state.
Exact diagonalization results are consistent with this quantum spin liquid over an extended region of the ground state phase diagram although potential quantum critical points within this region could indicate a richer phase structure.
Our work thus expands the physics of quantum spin ice in an experimentally motivated geometry, showing how the emergent QED can be coupled to dynamical matter at zero temperature.
\end{abstract}
\date{\today}
\maketitle

\section{Introduction}

Spin ice materials include several rare-earth pyrochlore oxides, modeled at the classical level by Ising spins on the pyrochlore lattice \cite{harris1997,udagawa2021}. Their ground states are extensively degenerate and described by an emergent Coulomb phase, whose gauge-charge excitations are the famous magnetic monopoles \cite{castelnovo2008a}. When Ising interactions are enhanced into an XXZ coupling between $S = 1/2$ spins, the quantum dynamics due to the transverse terms transforms the classical Coulomb phase into an emergent form of quantum electrodynamics (QED) \cite{hermele2004,banerjee2008,benton2012}. Over the years, quantum spin ice has offered a platform to explore variations of emergent QED whose speed of light and fine structure constant become system dependent \cite{pace2021a}, motivated by experiments on Pr$_2$Hf$_2$O$_7$ \cite{sibille2018a}, Ce$_2$Sn$_2$O$_7$ \cite{sibille2020a, poree2025} and Ce$_2$Zr$_2$O$_7$ \cite{smith2022}.

\begin{figure}
\centering\includegraphics[width=7cm]{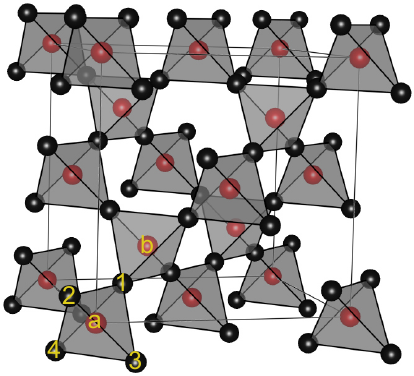}
\caption{The centered pyrochlore lattice with its 6 sublattices: $\{1,2,3,4\}$ for vertex sites, and $\{a,b\}$ for center sites.
The former make up a pyrochlore lattice while the latter occupy a diamond lattice.
}
\label{fig:latt}
\end{figure}

Metal-organic frameworks have recently put a spin on this physics. The centered pyrochlore lattice of Mn(ta)$_2$ \footnote{Where (ta) stands for the 1,2,3-triazolate ligand.} complements the traditional pyrochlore lattice with an additional spin in the center of all tetrahedra, increasing the number of sublattices from 4 to 6 [Fig.~\ref{fig:latt}] \cite{gandara12a,nutakki2023a}. 
In classical spin models, this central spin degree of freedom can be mapped onto a local gauge charge, with the ground state containing a finite density of them.  
For Ising spins, this opens a window of coupling parameters where monopoles are stabilized in the ground state, whereas for Heisenberg spins (a minimal model of the $S = 5/2$-compound Mn(ta)$_2$) the ground state charge density depends on the strength of exchange interactions~\cite{nutakki2023a,nutakki2023}.
 
However, the properties of quantum models on the centered pyrochlore lattice are completely unknown.
The goal of this paper is to ascertain whether a quantum model on the centered pyrochlore lattice can host a quantum spin liquid ground state, and if so, what kind of emergent lattice gauge theory it is described by.

On the pyrochlore lattice, the perturbative limit of the XXZ model hosts one of the most theoretically well-established and well-understood QSLs in the literature (quantum spin ice), due to the perturbative construction of its effective theory~\cite{hermele2004} and subsequent numerical evidence of a quantum liquid ground state~\cite{banerjee2008,shannon2012,huang2018}.
This provides the theoretical basis for understanding the properties of other quantum models~\cite{gingras2014, huang2014}, including those relevant to experimental candidate quantum spin-ice compounds.
Thus, we choose the XXZ model on the centered pyrochlore lattice as a theoretical model to study, which allows us to perturbatively derive an effective lattice gauge theory whose deconfined phase is a quantum spin liquid, distinct yet closely related to the rich physics studied on the pyrochlore lattice, and gives us the basic framework through which to understand the physics of quantum models on the centered pyrochlore lattice more generally.
By focusing on the parameter regime with ferromagnetic transverse interactions, meaning there is no sign problem, we set the stage for future numerical work to investigate further.

Whilst there are currently no known $S = 1/2$ compounds realizing a centered pyrochlore lattice, metal-organic frameworks are a versatile platform, due to the ability to substitute metals and ligands, so could be synthesized in the future.
Therefore, knowledge of what kinds of quantum spin liquid states are possible would be useful in understanding whether or not any material realizes such a state.
The contents of this article are as follows.
In section \ref{sec:model} we introduce the XXZ model on the centered pyrochlore lattice, and summarize its Ising physics \cite{nutakki2023}. In particular we present its $\mathbb{Z}_2$ classical spin liquid which corresponds to the region of the phase diagram we shall focus on.
In section \ref{sec:pert_theory}, we calculate in degenerate perturbation theory up to fourth order the first non-trivial term $B_{\hexagon}$ distinguishing our model from traditional quantum spin ice. $B_{\hexagon}$ induces a virtual quantum process which couples center spins around hexagonal loops of the lattice. 
In section \ref{sec:propHeff} we develop the corresponding quantum gauge field theory, in which the ground state is described by a \uo lattice gauge theory coupled to fermionic matter: The fermions carry an emergent electric charge, and the $B_{\hexagon}$ term of perturbation theory induces electric charge exchange around hexagons. Finally in section \ref{sec:ED} we present Exact Diagonalization
(ED) results which support a crystallization of the electric charges co-existing with a U(1) quantum spin liquid. We interpret these numerical results in the context of the perturbation and emergent gauge field theory derived previously, and link to the proximate quantum dimer model on the diamond lattice. We establish two points in the phase diagram as candidates for quantum criticality.

\section{Model}
\label{sec:model}
\begin{figure}[t]
\centering\includegraphics[width=\columnwidth]{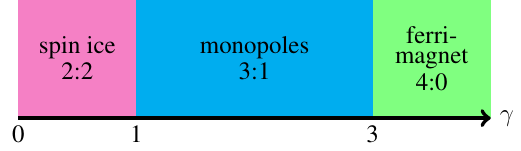}
\caption{
Ground-state phase diagram of the classical Ising model with $\gamma=J_1^z/J_2^z$ and $J_1^\perp=J_2^\perp=0$. Reproduced from \cite{nutakki2023}.
}
\label{fig:isingphd}
\end{figure}
Our work is inspired by the spin$-1/2$ XXZ model on pyrochlore, where it is well-established that the ground state in the perturbative $J^{\perp} \ll J^z$ limit is a $\mrm{U}(1)$ quantum spin liquid (QSL) \cite{hermele2004,banerjee2008,shannon2012,huang2018}. There, quantum fluctuations produce a superposition of the ground-state manifold of the Ising model. Similarly, on the centered pyrochlore lattice, we will use our knowledge of the Ising model~\cite{nutakki2023} as a starting point to understand how quantum fluctuations play a role. We consider the spin$-1/2$ XXZ model on the centered pyrochlore lattice
\begin{eqnarray}
	H &=& J_1^z \sum_{\langle ij \rangle} S_i^z S_j^z + J_1^{\perp} \sum_{\langle ij \rangle}\bigg(S_i^+ S_j^- + \mrm{h.c} \bigg)\nonumber\\
	&+& J_2^z \sum_{\langle\langle ij \rangle\rangle} S_i^z S_j^z + J_2^{\perp} \sum_{\langle \langle ij \rangle \rangle}\bigg(S_i^+ S_j^- + \mrm{h.c} \bigg),
    \label{eq:ham}
\end{eqnarray}
where $J_1^{z,\perp}$ terms act between nearest neighbors (center and vertex spins), $J_2^{z,\perp}$ between second neighbors (vertex spins) [Fig.~\ref{fig:latt}] and $S^\pm = S^x\pm i S^y$.
Whilst the $J_2$ XXZ anisotropy is a natural consequence of the pyrochlore symmetry \cite{curnoe2007,curnoe08a}, the anisotropy on the $J_1$ bond is not allowed by the cubic symmetry of the diamond lattice \cite{davier25} \footnote{A $J_1$ anisotropy could, however, appear spontaneously below a transition temperature selecting a preferred axis for the center spins.}
Rather than trying to describe a potential material, our approach here is to use the transverse couplings $J_{1,2}^{\perp}$ as a way to perturbatively include quantum dynamics to the classical model, in a controlled manner. 
Furthermore, at the SU(2) Heisenberg point $J_1^{\perp} = J_1^z/2$, $J_2^{\perp} = J_2^z/2$, the model is allowed by symmetry so a quantum spin liquid identified in the XXZ model could be a candidate for the ground state of a future $S=1/2$ Heisenberg compound.\\
\begin{figure}[t]
\centering\includegraphics[width=\linewidth]{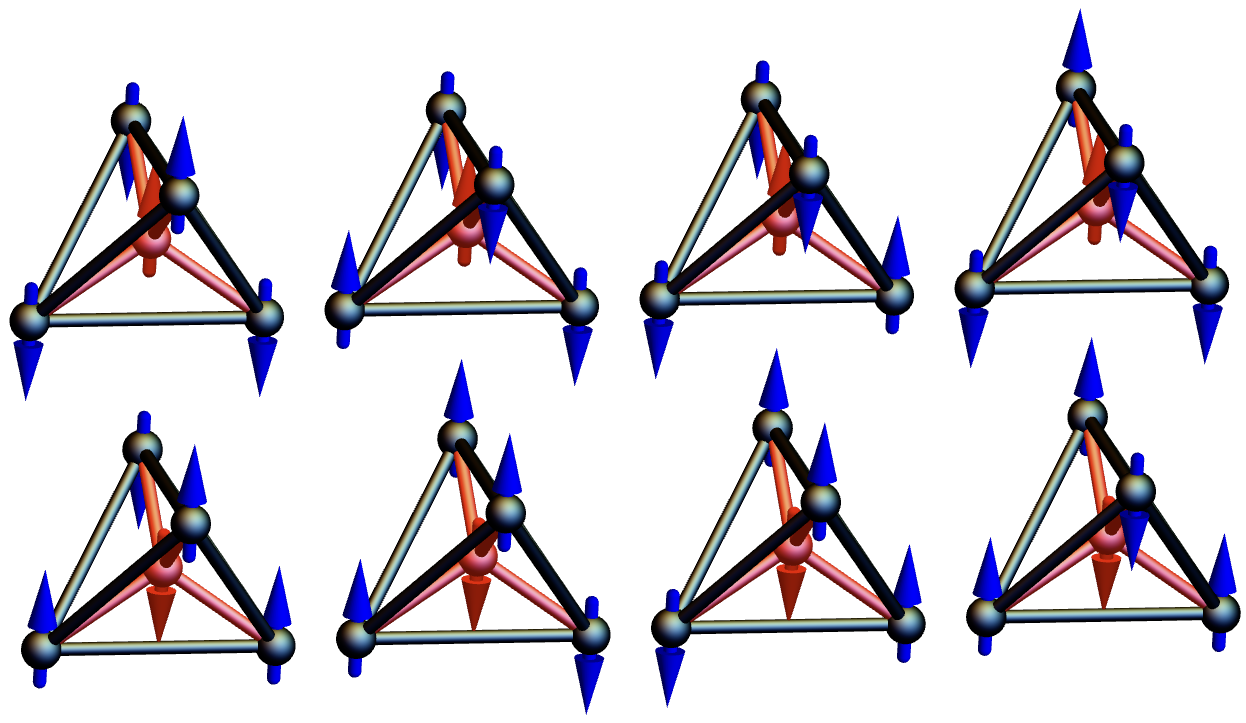}
\caption{
The eight allowed single-tetrahedron configurations of the Ising model ground state for $1<\gamma<3$.
The degenerate set of states which can be constructed from these configurations is a $\mathbb{Z}_2$ classical spin liquid~\cite{nutakki2023}. 
We refer to this ensemble of ground states as 3:1 states and $1<\gamma<3$ as the 3:1 regime.
Unlike 2:2 single-tetrahedron configurations one can obtain a valid 3:1 configuration from another by flipping a pair of majority vertex spins and the center spin, which is how the lower row is obtained from the upper row.
}
\label{fig:31regime}
\end{figure}

\subsection{Region of interest: the $\mathbb{Z}_2$ classical spin liquid}
\label{sec:Isingintro}

When $J_1^\perp=J_2^\perp=0$, $J_1^z, J_2^z>0$ and $\gamma=J_1^z/J_2^z$, the ground state of the centered pyrochlore Ising model is presented in Fig.~\ref{fig:isingphd}, reproduced from [\onlinecite{nutakki2023}]. For $0<\gamma<1$, it shares the same ground states as its pyrochlore counterpart, where a valid ground state configuration consists of two spins pointing up and two spins pointing down on the vertices of each tetrahedron; this is the well-known, extensively degenerate, spin-ice manifold \cite{harris1997}. In this ground state, the magnetization is the same for all (001) planes.
The center spins are completely decoupled from the vertex spins in this ground state. 
For $\gamma>3$, a ground state configuration has all vertex spins pointing in the same direction and all center spins pointing in the opposite direction, forming a trivial ferrimagnet.

However, the intermediate region, $1<\gamma<3$, supports a different ground state. Each tetrahedron has either 3-up/1-down or 3-down/1-up vertex spins, with the center spin aligned to the minority vertex spin, i.e. pointing respectively either down or up [Fig.~\ref{fig:31regime}]. This ground state is equivalent to a fully packed monopole liquid on the pyrochlore lattice~\cite{slobinsky2018,szabo2025}. This phase can be described as a {\zt} classical spinliquid due to its exponentially decaying correlations and {\zt} winding number; as opposed to the spin-ice manifold stabilized for $0<\gamma<1$, the magnetization of this {\zt} classical spin liquid can vary from one (001) plane to the other, but the parity of the magnetization of a (001) plane is constant across the system~\cite{nutakki2023}. However, despite its {\zt} nature, it is also possible to define an emergent, unique, electric field with finite divergence everywhere, which means that spins can be mapped to flux variables with local sources and sinks \cite{nutakki2023a,nutakki2023}. We refer to this region of the parameter space as the 3:1 regime, and a valid ground state spin configuration as a 3:1 state. 
\\

Previous works on the magnetization plateaux of spinels and the quantum dimer model on the diamond lattice ~\cite{bergman2006a,bergman2006,sikora2009,sikora2011} have considered a subset of the 3:1 regime with 3-up/1-down vertex spins on all tetrahedra (i.e. the four states in the second row of Fig.~\ref{fig:31regime} only). In both cases the magnetic field or dimer constraint induce a broken time-reversal symmetry in the spin language, resulting in an emergent U(1) gauge field even at the classical level. This is distinct from the 3:1 states of our model where both 3-up/1-down and 3-down/1-up states are allowed (all eight states of Fig.~\ref{fig:31regime}), and whose classical description corresponds to a \zt classical spin liquid \cite{nutakki2023} with a substantially larger ground-state degeneracy.
The quantum phase diagram of the complete 3:1 regime remains essentially unknown.
In addition, the $J_1^{\perp}$ interactions with the central spin induces an original path of quantum fluctuations that is absent from previous pyrochlore studies. 

\subsection{Hamiltonian symmetry}
\label{sec:Hamsym}

Let us first enumerate some of the symmetry properties of the Hamiltonian.
It is useful to rewrite the Hamiltonian as a sum over tetrahedra, $t$,
\begin{eqnarray}
	H = \sum_t H_t,
	\label{eq:hamt}
\end{eqnarray}
with
\begin{eqnarray}
	H_t &=& J_1^z \sum_{v=1}^4 S_{t,c}^z S_{t,v}^z + J_1^{\perp} \sum_{v=1}^4 \bigg(S_{t,c}^+ S_{t,v}^- + \mrm{h.c}\bigg)\nonumber\\ 
	&+&\frac{J_2^z}{2} \sum_{w\neq v}S_{t,v}^z S_{t,w}^z + J_2^{\perp} \sum_{w\neq v}S_{t,v}^+ S_{t,w}^-,
\end{eqnarray}
where $c$ labels center sites and $v,w$ label vertex sites of tetrahedron $t$.

Since the $J_1$ terms are bipartite in the sense that they only connect center to vertex spins, there is a pair of operators acting on center spins which map between different regions of the parameter space.
Consider a unitary operator,
\begin{eqnarray}
	U_c = \prod_{t} U_{t,c},
	\label{eq:Uc}
\end{eqnarray}
which acts only on center sites.
The only part of the Hamiltonian which does not commute with $U_c$ are the $J_1$ terms
\begin{eqnarray}
	H_{J_1} &=& J_1^z \sum_t S_{t,c}^z \sum_v S_{t,v}^z\nonumber\\
	&+& J_1^{\perp}\bigg(\sum_t S_{t,c}^+ \sum_v S_v^- + \sum_t S_{t,c}^- \sum_v S_v^+ \bigg).
\end{eqnarray}
An appropriate choice of $U_{t,c}$ will transform the center spins in such a way that $U_{t,c}^{\dagger}S^{\alpha}_{t,c}U_{t,c} = aS^{\alpha}_{t,c}$ and $a$ can be interpreted as a transformation of the exchange interaction, $J_1^{\alpha} \rightarrow a J_1^{\alpha}$.
Since vertex spins are coupled both to center spins and other vertex spins, an operator of the form Eq.~\eqref{eq:Uc} but acting on all vertex spins will not transform the exchange interactions in this way.

First, choosing $U_{t,c} = S_{t,c}^+ + S_{t,c}^-$, the inversion of the $z$-component of all center spins,
\begin{eqnarray}
	F_c = \prod_{t} \bigg(S_{t,c}^+ + S_{t,c}^-\bigg),
\end{eqnarray}
gives the mapping $(F_c)^{\dagger} H(J_1^z,J_2^z,J_1^{\perp},J_2^{\perp}) F_c = H(-J_1^z,J_2^z,J_1^{\perp},J_2^{\perp})$, that is, one can map from $J_1^z \rightarrow -J_1^z$ by inverting the $z$-component of all center spins.

Second, taking $U_{t,c} = r_{t,c}^z(\pi)$ as the single-site $\pi$ rotation operator in the local $z$ basis,
\begin{eqnarray}
	r_{t,c}^z(\pi) = \begin{pmatrix}
		-i & 0\\
		0 & i
	\end{pmatrix},
\end{eqnarray}
so that the $\pi$ rotation of all center spins about the $z$-axis,
\begin{eqnarray}
	R_c^z(\pi) = \prod_{t} r_{t,c}^z(\pi),
\end{eqnarray}
yields the mapping $(R_c^z(\pi))^{\dagger} H(J_1^z,J_2^z,J_1^{\perp},J_2^{\perp}) R_c^z(\pi) = H(J_1^z,J_2^z,-J_1^{\perp},J_2^{\perp})$, i.e rotation of all central spin components about the $z$-axis maps from $J_1^{\perp} \rightarrow -J_1^{\perp}$. Together, these two mappings are analogous to that in the classical Heisenberg model, where inverting all components of the central spins maps $J_1 \rightarrow -J_1$ [\onlinecite{nutakki2023a}].

In this paper, we largely present results for $J_1^z > 0$, $J_1^{\perp} < 0$, but they can be generalized to the opposite sign regions with a corresponding global inversion or rotation of the center spins.

\section{Perturbation Theory}
\label{sec:pert_theory}

We apply the Schrieffer-Wolff degenerate perturbation theory~\cite{schrieffer1966,bravyi2011} in the $J_1^{\perp},J_2^{\perp} \ll J_1^z,J_2^z$ limit, to understand the effect of quantum fluctuations to the manifold of 3:1 states. We refer to Ref.~[\onlinecite{slagle2017}] for a pedagogical introduction of the method and to Ref.~[\onlinecite{hermele2004}] for its application to pyrochlore. Following the rewriting of the Hamiltonian in terms of sums over tetrahedra [Eq.~(\ref{eq:hamt})], we now split it into longitudinal (Ising) and transverse (quantum fluctuation) parts,
\begin{eqnarray}
	\begin{split}
		H& = H^z + H^{\perp}, \\
		H^z &= J_1^z \sum_{\langle ij \rangle} S_i^z S_j^z + J_2^z \sum_{\langle\langle ij \rangle\rangle} S_i^z S_j^z  \\
		&= \frac{J_2^z}{2} \sum_t G_t^2 - \frac{N_t}{4}\bigg(\frac{(J_1^z)^2}{2J_2^z} + 2J_2^z\bigg)
		= \sum_t H^z_t + \mrm{const} \\[5pt]
		H^{\perp} &= J_1^{\perp} \sum_{\langle ij \rangle} \bigg(S_i^+ S_j^- + \mrm{h.c}\bigg)  + J_2^{\perp} \sum_{\langle \langle ij \rangle \rangle} \bigg( S_i^+ S_j^- + \mrm{h.c} \bigg)  \\
		&= J_1^{\perp}\sum_t \sum_{v=1}^4 \bigg( S_{t,c}^+ S_{t,v}^- + \mrm{h.c} \bigg)
		+ J_2^{\perp} \sum_t \sum_{w \neq v} S_{t,v}^+ S_{t,w}^-  \\
		&= H_{1}^{\perp} + H_{2}^{\perp},
	\end{split}
	\label{eq:H_pert}
\end{eqnarray}
where
\begin{eqnarray}
G_t = \eta_t \bigg(\gamma S_{t,c}^z + \sum_{v=1}^4 S_{t,v}^z \bigg),
\label{eq:Gdef}
\end{eqnarray}
with the convention that $\eta_t = +(-1)$ when the tetrahedron $t$ is on the $a(b)$ diamond sublattice. This convention will be justified when defining the effective Hamiltonian in section \ref{sec:propHeff}.

$H^z$ is the Ising model presented in section \ref{sec:Isingintro} \cite{nutakki2023}. Let us define the operator, $P$, which projects onto the ground state manifold of $H^z$, and
\begin{eqnarray}
	D = \frac{1-P}{E_0 - H^z},
\end{eqnarray}
which projects onto the manifold of excited states with a prefactor determined by the energy of the excited state relative to the ground state.
In the Schrieffer-Wolff formalism, our goal is to construct an effective Hamiltonian, $H_{p}$, which describes the effect of $H^{\perp}$ via virtual quantum processes acting within the degenerate ground state manifold, which we build up order by order.

For simplicity, we focus our analysis on $\gamma=2$ which is at the center of the classical 3:1 regime (see Table \ref{tab:1}). When $\gamma=2$, the ground state energy of $H_t^z$ is $E_0 = 0$, and the 4:0 and 2:2 states are degenerate, lowest-energy excitations for a tetrahedron. This way, virtual processes that leave the spin configuration unchanged become constants, and we can focus our analysis on the non-trivial off-diagonal terms which are essential for the description of an emergent gauge field theory.
We also assume that $J_1^{\perp},J_2^{\perp}$ are of the same order.

\begin{table}
	\centering
		\begin{tabular}{ccccc}
			\hline
			Abbreviation & State & $\abs{G_t}$ & $H^z_t(\gamma)$ & $H^z_t(\gamma = 2)$\\
			\hline
			\vspace{-3mm}
			& & & & \\
			\vspace{1mm}
			4:0 & $\ket{\Da \ua \ua \ua \ua}$& $\abs{-\frac{\gamma}{2}+2}$ & $\frac{J_2^z}{2}(-\frac{\gamma}{2}+2)^2$ & $\frac{J_2^z}{2}$\\
			\vspace{1mm}
			3:1 & $\ket{\Da \ua \ua \ua \da}$ & $\abs{-\frac{\gamma}{2}+1}$ & $\frac{J_2^z}{2}(-\frac{\gamma}{2}+1)^2$ & $0$\\
			\vspace{1mm}
			2:2& $\ket{\Da \ua \ua \da \da}$ & $\frac{\gamma}{2}$ & $\frac{J_2^z\gamma^2}{8}$ & $\frac{J_2^z}{2}$\\
			\vspace{1mm}
			3:1$^*$& $\ket{\Da \ua \da \da \da}$ & $\frac{\gamma}{2}+1$ & $\frac{J_2^z}{2}(\frac{\gamma}{2}+1)^2$ & $2J_2^z$\\
			\vspace{1mm}
			4:0$^*$& $\ket{\Da \da \da \da \da}$ & $\frac{\gamma}{2}+2$ & $\frac{J_2^z}{2}(\frac{\gamma}{2}+2)^2$ & $\frac{9J_2^z}{2}$\\
			\hline
		\end{tabular}
	\caption{\label{tab:1} 
	Eigenvalues of $\abs{G_t}$ and $H^z_t$ for all single tetrahedron states. A representative state is shown in the $S^z$ basis, with center spins indicated by a double arrow, vertex spins by a single arrow.
    The asterisk indicates that the center spin is parallel to the majority spin of the vertex spins, which results in a higher energy for $J_1^z>0$.}
\end{table}
\begin{figure*}
	\vspace{3mm}
	\begin{textblock}{3}(1.8,-0.1)
		{(a) \small $PH_1^{\perp}DH_1^{\perp}P$}
	\end{textblock}
	\begin{textblock}{3}(8,-0.1)
		{(b) \small $PH_2^{\perp}DH_2^{\perp}P$}
	\end{textblock}
	\begin{textblock}{3}(2,1.2)
		{\footnotesize $H_{t_0}^z=H_{t_2}^z=\frac{J_2^z}{2}$}
	\end{textblock}
	\begin{textblock}{3}(8,1.2)
		{\footnotesize $H_{t_1}^z=H_{t_2}^z=\frac{J_2^z}{2}$}
	\end{textblock}
	\includegraphics[width=7.5cm]{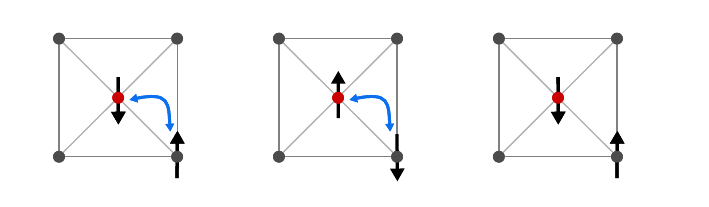}
	\qquad
	\includegraphics[width=7.5cm]{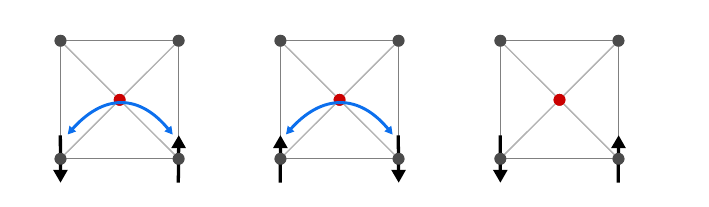}
	\begin{textblock}{1}(1.2,-0.5)
		{\small $0$}
	\end{textblock}
	\begin{textblock}{1}(1.8,-0.25)
		{\small $2$}
	\end{textblock}
	\begin{textblock}{1}(0.6,-0.25)
		{\small $1$}
	\end{textblock}
	\begin{textblock}{1}(7.35,-0.5)
		{\small $0$}
	\end{textblock}
	\begin{textblock}{1}(7.9,-0.25)
		{\small $2$}
	\end{textblock}
	\begin{textblock}{1}(6.75,-0.25)
		{\small $1$}
	\end{textblock}
	\vspace{3mm}
	\caption{
	Illustration of virtual processes contributing to the constant terms of Eq.~(\ref{eq:Hp2}) in second-order perturbation theory, reading from left (initial) to right (final). The initial and final states are always the same, while the intermediate (virtual) states bears two excited tetrahedra: (a) the one displayed here and the neighboring one connected to spin 2 (not shown), or (b) the two neighboring tetrahedra connected to spins 1 and 2. These virtual excitations are either 2:2 or 4:0 states, and are degenerate for $\gamma=2$, with energy $H_t^z=\frac{J_2^z}{2} \Rightarrow H^z=J_2^z$.
	}
	\label{fig:pert_2}
\end{figure*}

\subsection{Constraints on the virtual processes in a tetrahedron}
\label{sec:constraintvirtual}

The non-zero terms we obtain in perturbation theory will have the general form
\begin{equation}
    P A \dots B P,
\end{equation}
where we start with a ground state, apply some operators, $B,\dots,A$, to construct virtual processes, before returning to the ground state.
Therefore, the virtual processes we consider must map from a valid 3:1 state to another.
Here, we list the constraints this imposes at the level of a single tetrahedron.
In the following subsections, our methodology will be to make sure that the virtual processes respect these constraints on all tetrahedra in the system.
We encourage the reader to keep Table \ref{tab:1} in mind throughout the discussion.

For a given tetrahedron, single spin flips alone are trivially forbidden, as they necessarily induce an excitation. 
Similarly, flipping only the center and a vertex spin (e.g. via $J_1^\perp$) is also forbidden.
The simplest allowed virtual process is to flip two vertex spins, where one of these two vertex spins must be the minority spin of the configuration.
This is possible via $J_2^\perp$, $(J_1^\perp)^2$, $(J_2^\perp)^2$ ... terms.

One of the main differences with the traditional 2:2 regime of quantum spin ice~\cite{hermele2004}, or with the distinct 3:1 regime of spinels magnetization plateau with time-reversal broken symmetry~\cite{bergman2006a}, is that flipping two vertex spins pointing in the same direction is now possible; flipping the center spin together with two majority spins leaves the tetrahedron in its ground state (see~Fig.~\ref{fig:31regime}). 
This can be done for example via a $J_1^\perp J_2^\perp$ process, where the $J_2^\perp$ term acts on a neighboring tetrahedron. 
It is this move that is responsible for the \zt rather than \uo winding number in the classical spin liquid~\cite{nutakki2023}. 
We aim at reaching high enough perturbative order for such quantum dynamics to be accessible.

Finally, virtual processes flipping three vertex spins, or four spins in general, will always create excitations and are therefore forbidden for any tetrahedron.
Furthermore, while five-spin flip processes may be valid, they would require to go to seventh order in perturbation theory, which will not be considered in this paper.
\subsection{Second-order perturbation theory}

At second order (the lowest contributing order) the effective Hamiltonian is given by
\begin{eqnarray}
	\begin{split}
		H_{p}^{(2)} &= PH^{\perp}DH^{\perp}P\\
		&= PH^{\perp}_1DH^{\perp}_1P + PH^{\perp}_2 D H^{\perp}_2 P,
	\end{split}
	\label{eq:Hp2}
\end{eqnarray}
since $PH^{\perp}_1DH^{\perp}_2P$ terms are zero.
This is because $H^{\perp}_1 D H^{\perp}_2$ cannot connect two 3:1 states together, even if $J_1^{\perp}$ and $J_2^{\perp}$ terms act on the same tetrahedron and preserve its 3:1 configuration, they create an excitation on neighboring tetrahedra.

On the other hand, the two remaining terms of Eq.~(\ref{eq:Hp2}) map any ground state back to itself. 
They correspond to the two virtual processes shown in Fig.~\ref{fig:pert_2}, whose intermediate (virtual) states contain either 2:2 or 4:0 tetrahedral states. 
At $\gamma = 2$, the degeneracy of these virtual excitations simplifies the calculation of Eq.~(\ref{eq:Hp2}), which evaluates to a constant
\begin{eqnarray}
	\begin{split}
		PH^{\perp}_1DH^{\perp}_1P &= -\frac{(J_1^{\perp})^2}{J_2^z} \sum_t \sum_v \bigg(S_{t,c}^- S_{t,v}^+ S_{t,c}^+ S_{t,v}^- +\mrm{h.c}\bigg)\\
		&= -\frac{3N_t(J_1^{\perp})^2}{J_2^z}
	\end{split}
\end{eqnarray}
and
\begin{eqnarray}
		PH^{\perp}_2DH^{\perp}_2P &=& -\frac{(J_2^{\perp})^2}{J_2^z} \sum_t \sum_{v \neq w} S_{t,v}^- S_{t,w}^+ S_{t,v}^+ S_{t,w}^-\nonumber\\
		&=& -\frac{3N_t(J_2^{\perp})^2}{J_2^z},
\end{eqnarray}
where the factor of $3$ of each equation comes from the $3$ ways to get a virtual excitation. 
For a given 3:1 single-tetrahedron spin configuration, there are three majority vertex spins which can exchange either with the center spin of opposite direction ($PH^{\perp}_1DH^{\perp}_1P$ in Fig.~\ref{fig:pert_2}(a)), or with the minority vertex spin ($PH^{\perp}_2 DH^{\perp}_2P$ in Fig.~\ref{fig:pert_2}(b)).
Since constant terms apply equally to all ground states, they do not tell us anything about how the perturbation selects a specific ground state(s) and excitation structure out of this manifold, and are thus irrelevant for our discussion.

As a side comment, when $\gamma \neq 2$ (but ground states are still in the 3:1 manifold), these terms are no longer constant.
For $\gamma < 2$, 2:2 states are lower in energy than 4:0, so we expect the degeneracy of the 3:1 manifold to be broken by the second order term, with states where the perturbation can create the highest number of virtual 2:2 tetrahedra becoming lower in energy. 
Similarly, for $\gamma > 2$, states where the perturbation can create the highest number of virtual 4:0 states will have lower energy.
Whether this selects an ordered state or effectively reduces the dimension of the ground state manifold is not immediately clear.

\begin{figure*}
\centering\includegraphics[width=16cm]{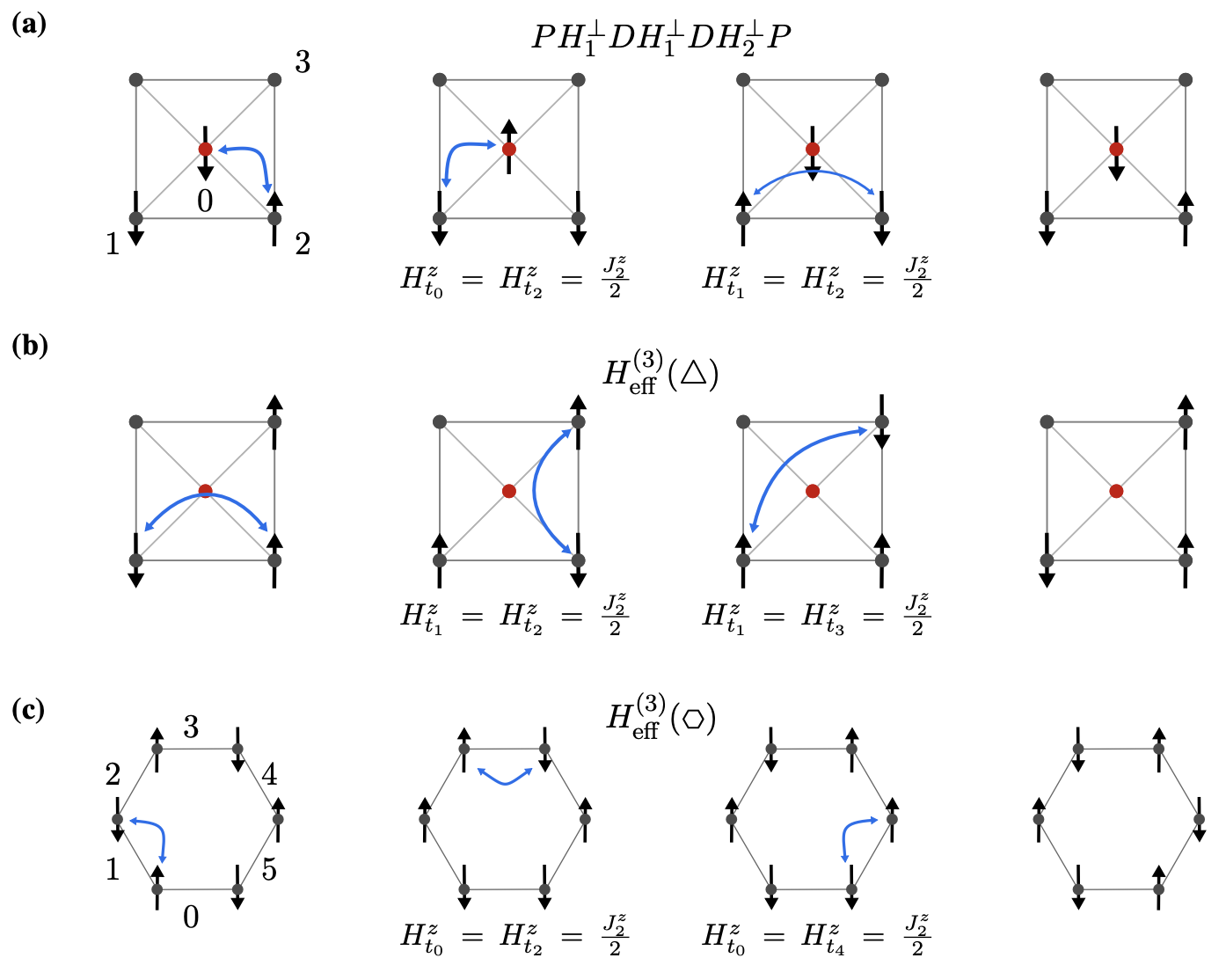}
\caption{
Processes contributing to third-order perturbation theory for $\gamma = 2$, with the single-tetrahedron energies of the intermediate (virtual) states shown below: $t_0$ is the displayed tetrahedron, $t_{i=\{1,2,3\}}$ are the neighboring tetrahedra connected via site $i$. (a) and (b) are triangular-hoping processes, respectively mediated by a center spin [Eq.~(\ref{eq:Hp3CPy1})], or not [Eq.~(\ref{eq:Hp3triang})]. (c) is the off-diagonal ring-exchange term, with  clockwise convention to label the sites and tetrahedra around an hexagonal loop. Only the edges of the tetrahedra making up the loop are shown.
}
\label{fig:pert_3}
\end{figure*}

\subsection{Third-order perturbation theory}

Closed loops become possible at third order, such as the well-known triangular-hopping and ring-exchange terms of the pyrochlore lattice \cite{hermele2004} thanks to the $J_2^\perp$ coupling
\begin{eqnarray}
	PH_2^{\perp}DH_2^{\perp}DH_2^{\perp}P = H_{p\;\triangle}^{(3)} + H_{p\; \hexagon}^{(3)}.
\end{eqnarray}
where, for $\gamma=2$,
\begin{widetext}
\begin{eqnarray}
H_{p\;\triangle}^{(3)} = \frac{(J_2^{\perp})^3}{(J_2^z)^2} \sum_t \sum_{u,v,w \in \triangle} \bigg( S_{t,w}^+ S_{t,u}^- S_{t,u}^+ S_{t,v}^- S_{t,v}^+ S_{t,w}^- +\mrm{h.c}\bigg)= \frac{6N_t(J_2^{\perp})^3}{(J_2^z)^2},
\label{eq:Hp3triang}
\end{eqnarray}
\begin{eqnarray}
H_{p\;\hexagon}^{(3)} = K_1 \sum_{\hexagon} \bigg(S_{v,0}^+ S_{v,1}^- S_{v,2}^+ S_{v,3}^- S_{v,4}^+ S_{v,5}^- +\mrm{h.c}\bigg)= K_1 \sum_{\hexagon} \bigg(A_{\hexagon} +A_{\hexagon}^{\dagger} \bigg)
\label{eq:Hp3ring}
\end{eqnarray}
with
\begin{eqnarray}
A_{\hexagon} = S_{v,0}^+ S_{v,1}^- S_{v,2}^+ S_{v,3}^- S_{v,4}^+ S_{v,5}^-
\quad\mathrm{and}\quad
K_1 = \dfrac{12(J_2^{\perp})^3}{(J_2^z)^2}.
\end{eqnarray}
\end{widetext}
The sum of Eq.~(\ref{eq:Hp3triang}) is over all vertex sites ${u \neq v \neq w}$ which make up a triangle on a single tetrahedron, as shown in Fig.~\ref{fig:pert_3}(b). 
As for Eq.~(\ref{eq:Hp3ring}), the vertex sites are labelled in a clockwise fashion around the hexagonal loop, as shown in Fig.~\ref{fig:pert_3}(c).
The prefactors again come from counting the number of processes which return to the original 3:1 state~\cite{hermele2004}.

On the centered pyrochlore lattice, additional virtual processes,
\begin{eqnarray}
2PH_1^{\perp}DH_1^{\perp}DH_2^{\perp}P + PH_1^{\perp}DH_2^{\perp}DH_1^{\perp}P,
\end{eqnarray}
are possible, mediated by the $J_1^{\perp}$ coupling with the center spin [Fig.~\ref{fig:pert_3}(a)].
The factor of 2 comes from the equivalence between $PH_1^{\perp}DH_1^{\perp}DH_2^{\perp}P$ and $PH_2^{\perp}DH_1^{\perp}DH_1^{\perp}P$. These terms, however, contribute only up to a constant
\begin{widetext}
\begin{eqnarray}
PH_1^{\perp}DH_1^{\perp}DH_2^{\perp}P = \frac{(J_1^{\perp})^2J_2^{\perp}}{(J_2^z)^2} \sum_t \sum_{v<w} \bigg(S_{t,c}^+S_{t,v}^-S_{t,c}^-S_{t,w}^+S_{t,w}^-S_{t,v}^+ + S_{t,c}^-S_{t,v}^+S_{t,c}^+S_{t,w}^-S_{t,w}^+S_{t,v}^- \bigg) = \frac{6N_t(J_1^{\perp})^2J_2^{\perp}}{(J_2^z)^2},
\label{eq:Hp3CPy1}\\
PH_1^{\perp}DH_2^{\perp}DH_1^{\perp}P = \frac{(J_1^{\perp})^2J_2^{\perp}}{(J_2^z)^2} \sum_t \sum_{v<w} \bigg(S_{t,c}^+S_{t,v}^-S_{t,v}^+S_{t,w}^-S_{t,w}^+S_{t,c}^- + S_{t,c}^- S_{t,v}^+ S_{t,v}^- S_{t,w}^+ S_{t,w}^- S_{t,c}^+ \bigg) = \frac{6N_t(J_1^{\perp})^2J_2^{\perp}}{(J_2^z)^2}.
\end{eqnarray}
As a result, the third-order perturbation term on the centered pyrochlore lattice is
\begin{eqnarray}
H_{p}^{(3)} = 2PH_1^{\perp}DH_1^{\perp}DH_2^{\perp}P + PH_1^{\perp}DH_2^{\perp}DH_1^{\perp}P + PH_2^{\perp}DH_2^{\perp}DH_2^{\perp}P = K_1 \sum_{\hexagon} \bigg(A_{\hexagon} +A_{\hexagon}^{\dagger}\bigg) \;+\;\mathrm{cst},
\label{eq:Hp3hex}
\end{eqnarray}
\end{widetext}
which is the same as on the pyrochlore up to a constant. 
Hence, we need to move to fourth-order perturbation to see if a distinct perturbative term appears.

\subsection{Fourth-order perturbation theory}

At fourth order, the expression of the effective Hamiltonian becomes formally more complex
\begin{eqnarray}
\label{eq:Hp4tot}
H_{p}^{(4)} 
&=& P H^{\perp}D H^{\perp} D H^{\perp} D H^{\perp} P\\
&-&\frac{1}{2} P H^{\perp}D H^{\perp} P H^{\perp} D^2 H^{\perp} P \nonumber\\
&-&\frac{1}{2} P H^{\perp}D^2 H^{\perp} P H^{\perp} D H^{\perp} P,\nonumber
\end{eqnarray}
where the second and third term of the right-hand side correspond to pairs of second-order virtual processes; the intermediate $P$ projector annihilates all excited states, enforcing to recover a 3:1 state after each $H^\perp D H^\perp$ or $H^\perp D^2 H^\perp$ processes. 
Since we have already established that second-order processes only contribute up to an irrelevant constant in the effective Hamiltonian for $\gamma=2$, we will not further consider them here. 
Instead we shall focus on the first term of Eq.~(\ref{eq:Hp4tot}) and more precisely on the off-diagonal terms it contains. 
In other words, the virtual processes between distinct 3:1 states which are possible with four $J^\perp$ quantum steps.

It is easy to see that, among these four steps, having a unique $J_1^\perp$ step is not possible as it would involve an odd number of vertex spins.
This would necessarily leave at least one tetrahedron with an odd number of flipped vertex spins, and thus induce an excitation (see Table \ref{tab:1}).
The same reasoning applies for three $J_1^\perp$ steps.

Based on section \ref{sec:constraintvirtual}, four $J_1^\perp$ steps can only produce a constant, because, in the absence of $J_2^\perp$ steps, we need to impose two $J_1^\perp$ steps per tetrahedron in order to remain in a 3:1 state. 
Since four $J_1^\perp$ steps are not enough to form a closed loop in the lattice (this would only be possible at sixth-order perturbation theory) the process maps a 3:1 state back to itself and therefore is a constant.
\begin{figure}[t]
\centering\includegraphics[width=\columnwidth]{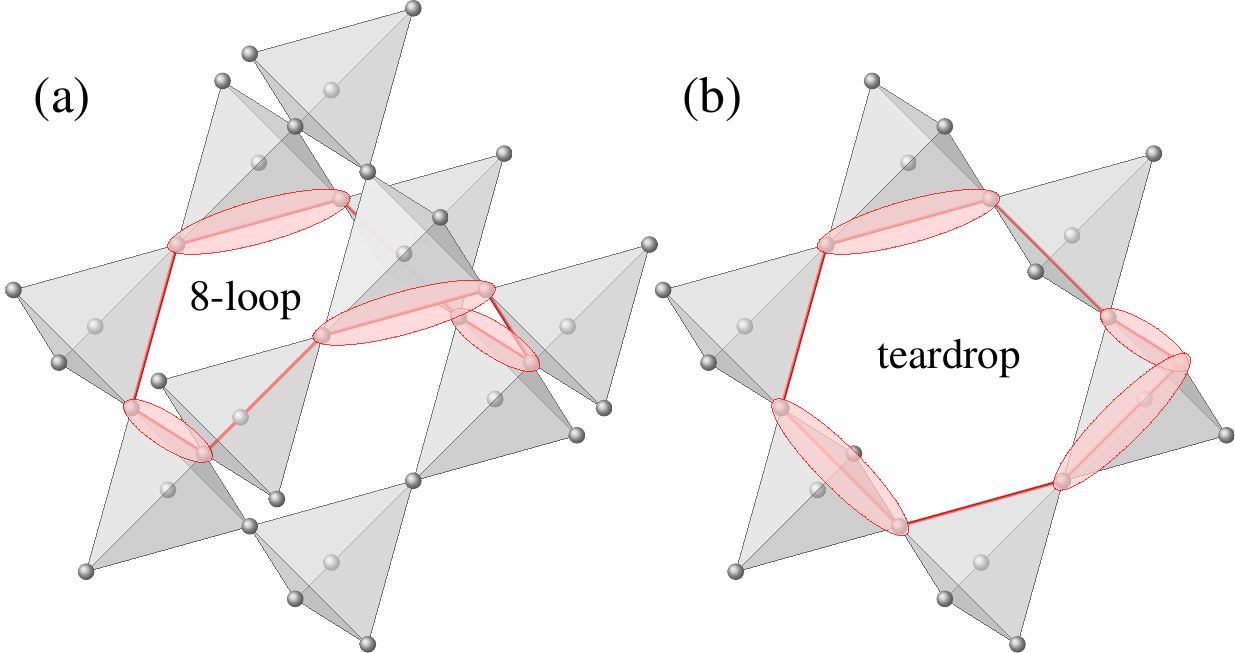}
\caption{
Two off-diagonal processes contributing to fourth-order perturbation theory with only $J_2^\perp$ dynamics: (a) the 8-site loop and (b) the ``teardrop'' \cite{Sanders2024}. Since one of the spin is flipped twice in (b), the teardrop is equivalent to an hexagonal loop. Ellipses represent bond operators used in the virtual process.
}
\label{fig:pert_4H2}
\end{figure}

As a result, Eq.~(\ref{eq:Hp4tot}) can be written as
\begin{eqnarray}
H_{p}^{(4)}	
&=& \sum_{ijkl =\pi\{1122\}} PH_i^{\perp} D H_j^{\perp} D H_k^{\perp} D H_l^{\perp} P \nonumber\\
&+& PH_2^{\perp} D H_2^{\perp} D H_2^{\perp} D H_2^{\perp} P
\label{eq:Hp4form}
\end{eqnarray}
up to a constant, where the first term contains the six permutations of the $H_1^{\perp}$, $H_2^{\perp}$ operators: \{1122, 1212, 1221, 2112, 2121, 2211\}.

Let us start with the second term, which is the same as on the pyrochlore \cite{Sanders2024} and contributes to two off-diagonal processes depicted in Fig.~\ref{fig:pert_4H2}. There is an 8-site loop dynamics
\begin{eqnarray}
H_{p\;\octagon}^{(4)} = -\frac{40(J_2^{\perp})^4}{(J_2^z)^3} \sum_{\octagon} \bigg( O_{\octagon} + O_{\octagon}^{\dagger} \bigg),
\label{eq:Hpoctagon}
\end{eqnarray}
where $O_{\octagon}\equiv S_{v,0}^+ S_{v,1}^- S_{v,2}^+ S_{v,3}^- S_{v,4}^+ S_{v,5}^-S_{v,6}^+ S_{v,7}^-$ whose sites are labelled clockwise around the octagon. 
There is also a ``teardrop'' process, as recently coined in Ref.~[\onlinecite{Sanders2024}]; it corresponds to a 7-site loop where one of the spins is flipped twice, resulting effectively in a rescaling of the prefactor of the ring term found at third order in perturbation
\begin{eqnarray}
H_{p\mrm{, teardrop}}^{(4)} = -\frac{10(J_2^{\perp})^4}{(J_2^z)^3} \sum_{\hexagon} \bigg( A_{\hexagon} + A_{\hexagon}^{\dagger} \bigg),
\label{eq:Hpteardrop}
\end{eqnarray}

\begin{figure}[t]
\centering\includegraphics[width=\columnwidth]{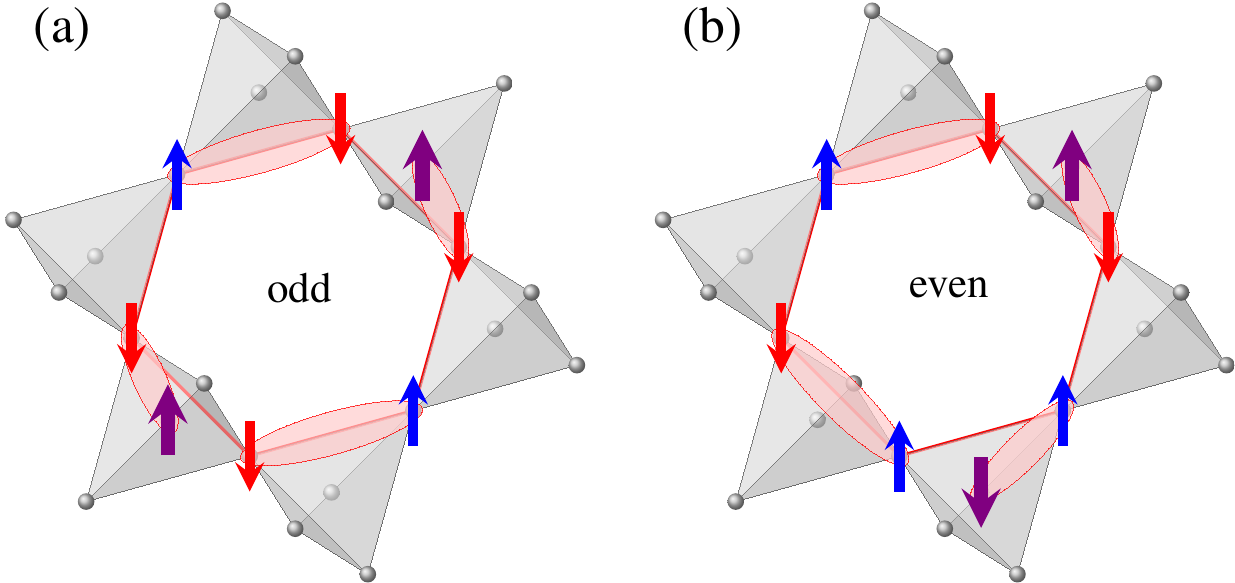}
\caption{
Two examples of off-diagonal processes contributing to fourth-order perturbation theory with two $J_1^\perp$ and two $J_2^\perp$ dynamics. The hexagonal loop of vertex spins is flipped, together with two center spins sitting at odd (a) and even (b) distance from each other along the loop. The important difference with the traditional ring exchange on pyrochlore is that the hexagonal loop is not made of $\uparrow\downarrow\uparrow\downarrow\uparrow\downarrow$ anymore. Instead, for each tetrahedron where a center spin (violet) is flipped, the loop flips two neighboring vertex spins pointing in the same $S^z$ direction. Even if the total magnetization is conserved because the process is made of $S^+S^-$ terms only, the magnetization of vertex spins is not conserved anymore when center spins are at odd distance from each other (a). Ellipses represent bond operators used in the virtual process.
}
\label{fig:pert_4H3}
\end{figure}

On the other hand, the first term of Eq.~(\ref{eq:Hp4form}) gives a non-trivial off-diagonal term, made of a six-site loop of vertex spins and 2 center spins. If the two center spins involved in the process are the same one, then we recover a teardrop-like process, which simply rescales the prefactor of Eq.~(\ref{eq:Hpteardrop}). 
But if the two center spins are distinct, then a new kind of resonating loop move appears, as illustrated in Fig.~\ref{fig:pert_4H3}. For reasons that will become clear in the next section, we shall refer to these updates as a ring-charge exchange, whose Hamiltonian is written as
%
%
\begin{eqnarray}
	\begin{gathered}
		H_{p\mrm{, ring-charge}}^{(4)} = -K_2 \sum_{\hexagon} \bigg( B_{\hexagon} + B_{\hexagon}^{\dagger} \bigg), \\
		K_2 = \frac{12(J_1^{\perp})^2(J_2^{\perp})^2}{(J_2^z)^3}, \:
		B_{\hexagon} = \sum_{i=0}^5\sum_{j=i+1}^5 B^{ij}_{\hexagon},\\
		B^{ij}_{\hexagon} = 
		\begin{cases}
		S_{c_i}^+ S_{c_j}^- S_{v_i}^- \dots S_{v_{j-1}}^+ S_{v_j}^+ \dots S_{v_{i+5}}^- \: \mrm{for} \: j-i \: \mrm{even}\\
		S_{c_i}^+ S_{c_j}^+ S_{v_i}^- \dots S_{v_{j-1}}^- S_{v_j}^- \dots S_{v_{i+5}}^- \: \mrm{for} \: j-i \: \mrm{odd}
		\end{cases}.
		\label{eq:Hp4}
	\end{gathered}
\end{eqnarray}
where the sums over $i$ and $j$ are made over all arrangements of the two center spins around the loop [Fig.~\ref{fig:pert_4H3}]. Explicitly writing out these terms for the two examples of Fig.~\ref{fig:pert_4H3},
\begin{eqnarray}
(a) B^{03}_{\hexagon} &= S_{c,0}^-S_{c,3}^- S_{v,0}^+ S_{v,1}^- S_{v,2}^+ S_{v,3}^+ S_{v,4}^- S_{v,5}^+,\\
(b) B^{02}_{\hexagon} &= S_{c,0}^-S_{c,2}^+ S_{v,0}^+ S_{v,1}^- S_{v,2}^- S_{v,3}^+ S_{v,4}^- S_{v,5}^+.
\end{eqnarray}
The factor of $12$ in the expression of $K_2$ comes from the six different arrangements of $H_1^{\perp}, H_2^{\perp}$ and the fact that for given $i,j$, which fixes the $J_1^{\perp}$ contribution, there is a choice of two positions on the hexagon for the first $J_2^{\perp}$ contribution, which then fixes the position of the second. 

It was thus necessary to go up to 4$^{\rm th}$ order in perturbation theory in order to find the first non-trivial term intrinsic to the \textit{centered} pyrochlore lattice.
Therefore, our effective Hamiltonian up to fourth order within the 3:1 manifold is
\begin{eqnarray}
	H_{p} = K_1 \sum_{\hexagon} \bigg(A_{\hexagon} +A_{\hexagon}^{\dagger} \bigg) -K_2 \sum_{\hexagon} \bigg( B_{\hexagon} + B_{\hexagon}^{\dagger} \bigg),
	\label{eq:Hp_spins}
\end{eqnarray}
where we did not include $H_{p\;\octagon}^{(4)}$ of Eq.~(\ref{eq:Hpoctagon}) because it sits on the traditional pyrochlore lattice, and is thus not intrinsic to the \textit{centered}-pyrochlore physics. We expect the 8-site ring exchange to have qualitatively the same effect as the traditional hexagonal ring exchange $H_{p\;\hexagon}$ on the emergent gauge theory (see next section).

\section{The lattice gauge theory $\mathsf{H_{\eff}}$}
\label{sec:propHeff}

\subsection{On the pyrochlore lattice}

\begin{figure}[b]
    \begin{minipage}{4cm}
        \includegraphics{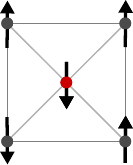}
    \end{minipage}
    \begin{minipage}{4cm}
        \includegraphics{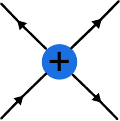}
    \end{minipage}
    \caption{(Left) A single tetrahedron 3:1 configuration of spins with arrows corresponding to $S_i^z = \pm 1/2$.
    (Right) The same spin configuration after mapping the vertex spins to directed rotors (electric field), $e_{rr'} = \pm 1/2$, and central spin to an effective (electric) charge.
    The rotors are defined on links of the diamond lattice which correspond to the sites of the pyrochlore lattice.
    The center site is on the $a$ sublattice; if on the $b$ sublattice, the sign of the charge is flipped as well as directions of the rotor variables.
    }
    \label{fig:spins_to_rotors}
\end{figure}

We now discuss how the perturbative Hamiltonian $H_{p}$ of Eq.~(\ref{eq:Hp_spins}) can be interpreted as an emergent lattice gauge theory, describing the low-energy properties of the system.
Let us start with the vertex spins on the pyrochlore lattice, as was done in Ref.~[\onlinecite{hermele2004}], that can be expressed in terms of directed conjugate quantum rotor variables,
\begin{eqnarray}
S_{v_i}^z \rightarrow \eta_r e_{rr'}, \qquad  S^{\pm}_{v_i} \rightarrow e^{\pm i\eta_r a_{rr'}},
\label{eq:Srotor}
\end{eqnarray}
where $r$ and $r'$ label diamond sites (i.e center sites) and $rr'$ the link joining the two sites (i.e a vertex site). As a reminder, $\eta_r = +(-1)$ when the diamond site $r$ is on the $a(b)$ sublattice.
$a_{rr'}$ and $e_{rr'}$ are conjugate variables, $[a_{rr'},e_{rr'}] = i$, and can be interpreted as a vector potential and electric field respectively. 
This mapping is illustrated in Fig.~\ref{fig:spins_to_rotors}.
The eigenvalues of $e_{rr'}$ are the set of all half-integers, whereas those of $a_{rr'} \in [0,2\pi)$. The mapping of Eq.~(\ref{eq:Srotor}) is exact provided the hard-core constraint $e_{rr'} \in \{-1/2,1/2\}$ is enforced. Since this mapping only applies to vertex spins on the pyrochlore lattice, we can only rewrite the effective Hamiltonian up to third order for the time being \cite{hermele2004}
\begin{eqnarray}
H_{\eff}^{(3)} = U_e \sum_{\langle rr' \rangle} e_{rr'}^2 +2K_1\sum_{\hexagon}	\cos(\mrm{curl}\:a)_{\hexagon},
\label{eq:Heff3}
\end{eqnarray}
where
\begin{eqnarray}
	(\mrm{curl}\:a)_{\hexagon} = \sum_{i=0}^5 a_{r_ir_{i+1}},
\end{eqnarray}
with center and vertex sites labelled clockwise around a hexagonal loop. 
The hard-core constraint on the $e_{rr'}$ is enforced by the first term of Eq.~(\ref{eq:Heff3}), taking the limit $(U_e/K_1) \rightarrow \infty$. Hamiltonian (\ref{eq:Heff3}) is invariant under gauge transformations
\begin{equation}
a_{rr'} \rightarrow a_{rr'} + \chi_{r'} - \chi_{r},\label{eq:gaugetrans}
\end{equation}
with $a_{rr'} \in [0,2\pi)$, making it a compact \uo lattice gauge theory.
Thus $a_{rr'}$ and $e_{rr'}$ can be interpreted as a vector potential and electric field respectively~\cite{hermele2004}. 
The lattice divergence of $e_{rr'}$,
\begin{eqnarray}
\Gamma_r = \mrm{div}(e)_r = \sum_{r' \in \mrm{nn}} e_{rr'},
\label{eq:Gdive}
\end{eqnarray}
generates the local \uo transformation of Eq.~(\ref{eq:gaugetrans}), and commutes with the Hamiltonian, $[H_{\eff}^{(3)},\Gamma_r]=0$. The sum in Eq.~(\ref{eq:Gdive}) runs over the four nearest neighbors sites $r'$ of $r$. Numerical simulations on the pyrochlore lattice~\cite{shannon2012} showed that the corresponding (non-overlapping) quantum dimer model to $H_{\eff}^{(3)}$ is a quantum liquid, which means that $H_{\eff}^{(3)}$ is deconfined for all $U_e/K_1$ in the $\mrm{div}(e)_r=0$ sector of the theory.
\\
\subsection{On the centered pyrochlore lattice}
However, in contrast to the pyrochlore, our model bears additional degrees of freedom, the center spins, which appear in the Ising ground state constraint $G_r =\gamma \eta_r S^z_r +\Gamma_r = 0$ of Eq.~(\ref{eq:Gdef}).
Therefore we are interested in the properties of the theory in a different gauge sector to the pyrochlore case, where it is not known if it will be deconfined in the $(U_e/K_1)\rightarrow \infty$ limit.

We can rewrite the constraint to the physical gauge sector by considering the center spins as static charges,
\begin{eqnarray}
\mrm{div}(e)_r = -2 \eta_r S^z_r \equiv q_r,
\label{eq:diveq}
\end{eqnarray}
where $q_r \in \{-1,1\}$ for $\gamma = 2$. 
Since $q_r$ does not appear explicitly in $H_{\eff}^{(3)}$, the charge configuration is an external parameter at third order in perturbation theory. 
Hence, one needs to move to fourth order to obtain non-zero matrix elements between different charge configurations.

\begin{figure}[t]
	\centering
	\begin{minipage}{8cm}
	\begin{tikzpicture}
		\node (figure1) at (0,0) {\includegraphics[width=\textwidth]{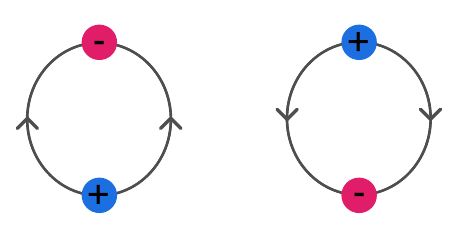}};
		\draw[->,black,ultra thick](-0.4,0) -- (0.4,0);
		\node (text1) at (0,0.5) {$\tilde{B}^{ij}_{\hexagon}$};
		\node (text2) at (-2.3,-2) {$r_i$};
		\node (text3) at (-2.3,1.9) {$r_j$};
		\node (text4) at (2.1,-2) {$r_i$};
		\node (text5) at (2.1,1.9) {$r_j$};
	\end{tikzpicture}
	\end{minipage}
	\caption{Schematic representation of the action of the $\tilde{B}^{ij}_{\hexagon}$ operator. The charges represent the fermionic spinons and the grey lines the electric field, $e_{rr'}$. The operators act on length-6 hexagonal loops.} 
	\label{fig:K2}
\end{figure}
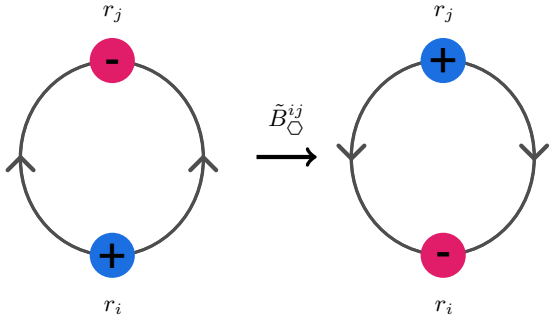

At fourth order, we map the center spins to Abrikosov fermions~\cite{savary2017}, where the mapping depends on the sublattice $\{a,b\}$ of the spins,
\begin{eqnarray}
	\begin{gathered}
		S_r^z = -\frac{\eta_r}{2} \bigg(f_{r+}^{\dag}f_{r+} - f_{r-}^{\dag}f_{r-}\bigg),\\
		S_r^+ = \begin{cases}
			f_{r-}^{\dagger}f_{r+}, \quad r \in a\\
			f_{r+}^{\dagger}f_{r-}, \quad r \in b
		\end{cases},\\
		S_r^- = \begin{cases}
			f_{r+}^{\dagger}f_{r-}, \quad r \in a\\
			f_{r-}^{\dagger}f_{r+}, \quad r \in b
		\end{cases}.
	\end{gathered}
	\label{eq:s_fermions}
\end{eqnarray}
We label the fermion flavour by an electric charge $+,-$, which gives
\begin{eqnarray}
G_r = n_{r-} -  n_{r+} + \mrm{div}(e)_r,
\label{eq:gauss_law}
\end{eqnarray}
where we have introduced the number operator $n_{r\pm} = f_{r\pm}^{\dagger}f_{r\pm}$ on all center sites (a.k.a. diamond sites). Hence, in the $G_r = 0$ sector imposed in the 3:1 regime [Eq.~(\ref{eq:Gdef})], we get
\begin{eqnarray}
\mrm{div}(e)_r =  n_{r+} - n_{r-},
\label{eq:diven}
\end{eqnarray}
This is an exact mapping provided the constraint $ n_{r+} + n_{r-} = 1$ is respected, i.e there is one fermion per center site. This will be enforced by adding the following term to the Hamiltonian
\begin{eqnarray}
 U_f \sum_r \bigg((n_{r+} - \frac{1}{2})(n_{r-} - \frac{1}{2})\bigg),
\end{eqnarray}
and taking the limit $U_f/K_2\rightarrow \infty$. In terms of Abrikosov fermions, the fourth-order operator of Eq.~(\ref{eq:Hp_spins}) becomes
\begin{widetext}
\begin{eqnarray}
	B^{ij}_{\hexagon} = \begin{cases}
		f_{r_i-}^{\dag}f_{r_i+}f_{r_j+}^{\dag}f_{r_j-}e^{i(\mrm{curl}^* a)_{ij}},\: \mrm{for} \: j-i \: \mrm{even}, \: r_i,r_j \in a\\
		f_{r_i+}^{\dag}f_{r_i-}f_{r_j-}^{\dag}f_{r_j+}e^{-i(\mrm{curl}^* a)_{ij}},\: \mrm{for} \: j-i \: \mrm{even}, \: r_i,r_j \in b\\
		f_{r_i-}^{\dag}f_{r_i+}f_{r_j+}^{\dag}f_{r_j-}e^{i(\mrm{curl}^* a)_{ij}},\: \mrm{for} \: j-i \: \mrm{odd}, \: r_i \in a, \: r_j \in b\\
		f_{r_i+}^{\dag}f_{r_i-}f_{r_j-}^{\dag}f_{r_j+}e^{-i(\mrm{curl}^* a)_{ij}},\: \mrm{for} \: j-i \: \mrm{odd}, \: r_i \in b, \: r_j \in a
		\end{cases}
	\label{eq:B_ij}
\end{eqnarray}
\end{widetext}
where a modified definition of the lattice $\mrm{curl}$ is required,
\begin{eqnarray}
(\mrm{curl}^* a)_{ij} = -\sum_{i'=i}^{j-1} a_{r_{i'}r_{i'+1}} + \sum_{i'=j}^{i-1} a_{r_{i'}r_{i'+1}}.
\label{eq:newcurl}
\end{eqnarray}
Since fermions on center spins bear an effective electric charge, Eq.~(\ref{eq:diven}) is not a zero-divergence condition anymore (as opposed to the traditional pyrochlore QSL \cite{hermele2004}). Within the emergent lattice gauge theory of our model, the presence of charges allows for two distinct electric field lines to be created at a $+$ fermion and to vanish in a $-$ fermion on the other side of a hexagonal loop.
This is illustrated in Fig.~\ref{fig:K2} and is the reason for the modified curl of Eq.~(\ref{eq:newcurl}) where each sum corresponds to a different electric field line.

At the microscopic level, in Fig.~\ref{fig:pert_4H3}(a), the two center spins involved in the virtual process belong to two distinct types of tetrahedra; their respective $\eta_r$ terms are thus opposite from each other, resulting in two opposite gauge charges since the two spins point in the same direction. Conversely, in Fig.~\ref{fig:pert_4H3}(b), the two center spins involved in the virtual process belong to the same type of tetrahedra; their $\eta_r$ terms are thus the same, which is why the two spins point in opposite direction in order to correspond to two opposite gauge charges. Hence, the virtual process $B_{\hexagon}$ respects charge conservation.
The fourth order part of the effective Hamiltonian can be written compactly as
\begin{eqnarray}
	\begin{gathered}
		H_{p}^{(4)} = - K_2 \sum_{\hexagon} \sum_{i=0}^5 \sum_{j = i+1}^5 \bigg(\tilde{B}^{ij}_{\hexagon} + \mrm{h.c.}\bigg),\\
		\tilde{B}^{ij}_{\hexagon} =
		 f_{r_i-}^{\dagger}f_{r_i+}f_{r_j+}^{\dagger}f_{r_j-}e^{i(\mrm{curl}^* a)_{ij}},\\
	\end{gathered}
\end{eqnarray}
where $\tilde{B}^{ij}_{\hexagon}$ hops a $+$ fermion from site $i$ to $j$ and a $-$ fermion from site $j$ to $i$, flipping the electric field on the two paths between them, thereby leaving $G_r$ invariant [Fig.~\ref{fig:K2}]. In the fermion and rotor language, the effective Hamiltonian (\ref{eq:Hp_spins}) thus becomes
\begin{widetext}
\begin{eqnarray}
	H_{\eff}^{(4)} &=& U_e \sum_{\langle rr' \rangle} e_{rr'}^2 + U_f \sum_r \bigg((n_{r+} - \frac{1}{2})(n_{r-} - \frac{1}{2})\bigg)
	+2K_1\sum_{\hexagon}\cos(\mrm{curl}\:a)_{\hexagon}\nonumber\\
	&-&K_2\sum_{\hexagon}\sum_{i=0,j=i+1}^5\bigg(f_{r_i-}^{\dagger}f_{r_i+}f_{r_j+}^{\dagger}f_{r_j-}e^{i(\mrm{curl}^* a)_{ij}} + \mrm{h.c} \bigg),
\label{eq:H_rotor4}
\end{eqnarray}
\end{widetext}
which corresponds to the physical $H_{p}$ in the $U_e, U_f \rightarrow \infty$ limit and $G_r = 0$ gauge sector, which enforces the appropriate constraints on the rotors and fermions.
By construction, $[H_{\eff}^{(4)},G_r]=0$, which is a \uo local symmetry, with the gauge transformation $e^{-i \chi_r G_r} O_r e^{i\chi_r G_r}$, where $O_r$ is a single-variable operator in $H_{\eff}^{(4)}$ associated with the site $r$, resulting in
\begin{eqnarray}
\begin{gathered}
a_{r_ir_j} \rightarrow a_{r_ir_j} + \chi_{r_j} - \chi_{r_i},\\
f_{r_i+} \rightarrow f_{r_i+}e^{-i\chi_{r_i}}, \:
f_{r_i-} \rightarrow f_{r_i-}e^{i\chi_{r_i}},
\label{eq:u1gauge}
\end{gathered}
\end{eqnarray}
for $\chi_r \in [0,2\pi)$.
Thus the effective theory is a frustrated compact \uo lattice gauge theory with fermionic matter.\\
Ref.~[\onlinecite{hermele2004}] discussed the topological properties of the \uo QSL on the pyrochlore in terms of the electric flux
\begin{eqnarray}
	\Phi_k^E = \sum_{\langle rr' \rangle \in \mrm{plane}} e_{rr'},
\end{eqnarray}
where the sum is over all links which pierce a plane oriented perpendicular to the cubic axis, $k \in \{x,y,z\}$, which is analogous to the winding number in classical spin ice. However, this quantity is not conserved by $H_{\eff}^{(4)}$ since
\begin{eqnarray}
	[\Phi_k^E,\tilde{B}^{ij}_{\hexagon}]= -2\tilde{B}^{ij}_{\hexagon}, \qquad
	[\Phi_k^E,(\tilde{B}^{ij}_{\hexagon})^{\dagger}]= (2\tilde{B}^{ij}_{\hexagon})^{\dagger},
\end{eqnarray}
where $i,j$ are such that the hexagonal loop is intersected by the plane.
This is the same algebra as raising and lowering operators, where $(\tilde{B}^{ij}_{\hexagon})^{\dagger}$ and $\tilde{B}^{ij}_{\hexagon}$ raise and lower the eigenvalue of $\Phi_k^E$ by 2 respectively.
The $\tilde{B}^{ij(\dagger)}_{\hexagon}$ are analogous to the moves in the classical Ising model ground state \cite{nutakki2023} which change the winding number by $\pm 2$, giving rise to a \zt winding number.
In the quantum effective model this property of $\Phi_k^E$ is a manifestation of the fact that sectors with even or odd numbers of $+1/2$ center spins have internal non-zero matrix elements but are mutually disconnected.
In the charge language this translates to the fact that the difference in population of the charge species, i.e the total charge,
\begin{eqnarray}
	N_q = N_+ - N_- = \sum_{r} \bigg( n_{r+} - n_{r-} \bigg),
\end{eqnarray}
is conserved by Hamiltonian (\ref{eq:H_rotor4}). 
In other words, while we have a local \uo gauge transformation [Eq.~(\ref{eq:u1gauge})] in the emergent lattice gauge theory of Hamiltonian (\ref{eq:H_rotor4}) derived from the perturbative action of the $S_i^{x,y}$ spin components, we can define a \zt conserved quantity based on the $S_j^z$ spin components, when $j$ is restricted to vertex spins only.

\subsection{Comparison of gauge theories}
Let us pause and discuss the similarities and differences between the effective lattice gauge theory (Eq.~(\ref{eq:H_rotor4})) we have derived on the centered pyrochlore lattice and that on the pyrochlore lattice (Eq.~(\ref{eq:Heff3})), as well as comparing to quantum electrodynamics in (3+1)D. 

The deconfined phase of the \uo lattice gauge theory described by $H_{\mathrm{eff}}^{(3)}$ has a gapless photon mode, and magnetic monopoles (not be confused with the magnetic monopoles of a dipolar/Ising model, these are alternatively called visons in the literature) associated with the $K_1$ term (consider defining a magnetic field variable $b \propto (\mathrm{curl} \, a)_{\hexagon}$), which are gapped~\cite{hermele2004}.
The gauge sector corresponding to the spin model ground state is where $\mathrm{div}(e)_r = 0$, i.e there are no electric charges (spinons).
Whilst $H_{\mathrm{eff}}^{(3)}$ is derived for the ground state, one can add spinons ``by hand", where the degenerate manifold has a single defect tetrahedron with $\sum_v S_{t,v}^z = 1 \rightarrow \mathrm{div}(e)_r  = \eta_r $, at an energy cost $\sim J_2^z$.
Therefore at finite temperature the system's properties are governed by fluctuations of spinons and magnetic monopoles coupled with the gauge field.

On the centered pyrochlore lattice, enforcing the single-spinon per site constraint, the ground state must contain exactly $N_t$ spinons.
Discarding $K_2$ (restricting to third order), there is no coupling between spinons and gauge fields and the spinons are static, defining a specific charge sector.
The theory is identical to that on the pyrochlore, but in a different charge sector, such that in the deconfined phase there are also gapless photons and gapped magnetic monopoles.
Going up to fourth order, the $K_2$ term introduces interactions between spinons, which allows for fluctuations in the spinon/charge density, in other words, dynamical matter (see Fig.~\ref{fig:K2}).
Thus the ground state on the centered pyrochlore now contains dynamical spinons.
Furthermore, the vector potential/magnetic field also interacts with the spinons via the $K_2$ term which could modify the magnetic monopole and photon excitations.
In both cases, additional spinons can be added to the system as excitations, either via flipping a center spin with energy cost $\sim J_1^z$ to create an opposite charge spinon-antispinon pair, or a vertex spin with energy cost $\sim J_2^z$ creating a single spinon.

It is also useful to compare $H_{\mathrm{eff}}^{(4)}$ to a compact, non-relativistic \uo lattice quantum electrodynamics with two flavors of massless fermion~\cite{kogut1975,xu2019},
\begin{equation}
    H_{\mathrm{QED}} = H_{\mathrm{eff}}^{(3)} + \sum_{\alpha = \pm} \sum_{\langle rr' \rangle} \bigg(f^{\dagger}_{r\alpha} e^{ia_{rr'}} f_{r' \alpha} + \mathrm{h.c} \bigg)
    \label{eq:qed}
\end{equation}
and Gauss law
\begin{equation}
    G_r = \mathrm{div}(e)_r + \sum_{\alpha} n_{\alpha} = 0.
\end{equation}
Firstly, the spinons in $H_{\mathrm{eff}}^{(4)}$ are also massless, since we are interested in the limit $U_f \rightarrow \infty$, enforcing one spinon per site, so this term only contributes a constant to the Hamiltonian.
Secondly, the two flavors of fermions present in both theories behave differently with respect to the Gauss law, in quantum electrodynamics electrons with spin up or down have the same, rather than opposite electric charge.
Finally, the coupling of the gauge field to the fermions in $H_{\mathrm{eff}}^{(4)}$ is more complicated than the minimal coupling of free fermions in Eq.~(\ref{eq:qed}), interacting with both the other flavor of fermion and a loop of gauge fields.
We note that our choice of mapping for the center spins is not the only one, other choices could result in different effective lattice gauge theories with bosonic or fermionic degrees of freedom coupled to the gauge fields.

Returning to the ground state of the XXZ model on the centered pyrochlore, the question going forward is if Eq.~\eqref{eq:H_rotor4} is deconfined in the physical limit and gauge sector. 
If so, this would correspond to a \uo QSL, different from the one on pyrochlore. Another possibility is for single fermions (spinons) to be deconfined but for spinon pairs to condense, which would lead to a \zt QSL via the Higgs mechanism~\cite{savary2017}. Also pertinent, is whether the $\gamma = 2$ point, where Eq.~\eqref{eq:H_rotor4} was derived, is a special point in parameter space or whether the low energy properties of the model in a broader region of the parameter space can be understood in the context of the lattice gauge theory. In the following sections, we attempt to gain some insight using exact diagonalization.

\begin{figure}
	\centering
	\begin{minipage}{6cm}
	\centering
	\vspace{0pt}
	\includegraphics[width=\textwidth]{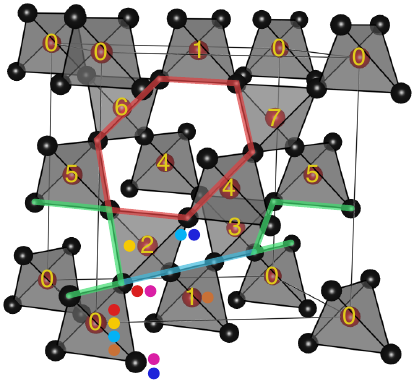}
	\end{minipage}
	\begin{minipage}{2cm}
	\centering
	\vspace{0pt}
	\begin{tabular}{cc}
		\hline
		Neighbors & Color\\
		\hline
		1st & \begin{tikzpicture}[baseline=-0.7ex]
			\definecolor{c}{rgb}{2.18, 0.31, 0.31}
			\draw[color=c, fill=c] (0, 0) circle (.18cm);
		\end{tikzpicture}\\
		2nd & \begin{tikzpicture}[baseline=-0.7ex]
			\definecolor{c}{rgb}{2.18, 0.31, 1.65}
			\draw[color=c, fill=c] (0, 0) circle (.18cm);
		\end{tikzpicture}\\
		3rd & \begin{tikzpicture}[baseline=-0.7ex]
			\definecolor{c}{rgb}{2.45, 2.01, 0.04}
			\draw[color=c, fill=c] (0, 0) circle (.18cm);
		\end{tikzpicture}\\
		4th & \begin{tikzpicture}[baseline=-0.7ex]
			\definecolor{c}{rgb}{0.04, 1.78, 2.45}
			\draw[color=c, fill=c] (0, 0) circle (.18cm);
		\end{tikzpicture}\\
		5th & \begin{tikzpicture}[baseline=-0.7ex]
			\definecolor{c}{rgb}{0.31, 0.36, 2.18}
			\draw[color=c, fill=c] (0, 0) circle (.18cm);
		\end{tikzpicture}\\
        6th & \begin{tikzpicture}[baseline=-0.7ex]
			\definecolor{c}{rgb}{200, 113, 55}
			\draw[color=brown, fill=brown] (0, 0) circle (.18cm);
		\end{tikzpicture}\\
		\hline
	\end{tabular}
	\end{minipage}
	\caption{The 24 site cluster used in exact diagonalization. The tetrahedra are numbered to show the periodic boundaries. The green and cyan highlighted paths are non-winding and winding loops of length-4 respectively, whereas the red path is the length-6 hexagonal loop. The six $(i,j)$ neighbor pairs with distinct distances between the sites are shown with matching dots, with the sorted list in order of distances shown in the table.}
	\label{fig:ed_24}
\end{figure}
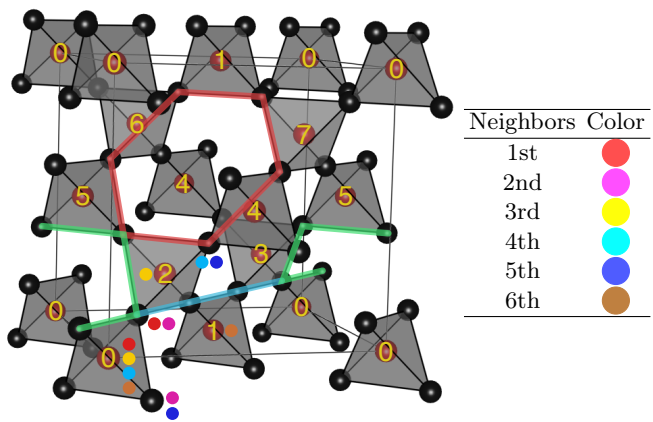

\begin{figure*}
	\centering
	\includegraphics[width=16cm]{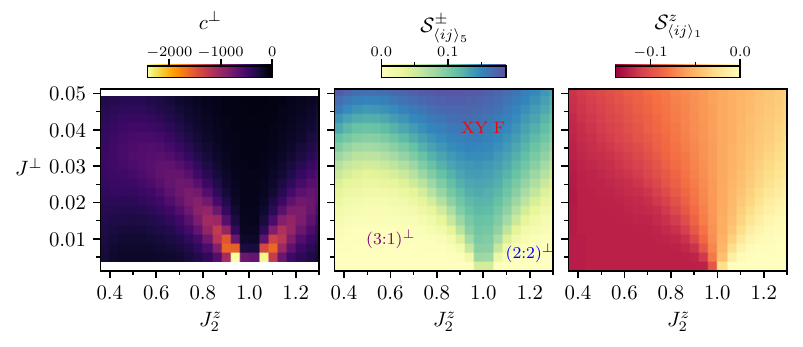}
	\caption{
	Ground state properties for ${{J_1^z = 1, J_1^{\perp} = J_2^{\perp} = -J^{\perp}/\sqrt{2}}}$ in the ${{m_z = 0}}$ sector obtained from ED for $N=24$. (Left) $c^{\perp}$ computed using finite differences, which is large in magnitude along two paths in the parameter space which meet at ${{J_2^z = 1}}$. (Middle) The $\mathcal{S}_{ij}^{\pm}$ correlator measured between 5$^{\rm th}$ nearest neighbors (the furthest distance between vertex spins on the $N=24$ cluster). For $J^{\perp}$ sufficiently large these correlations are large and positive, indicating the transition to a long-range XY ferromagnet. (Right) The $\mathcal{S}_{ij}^z$ correlator between nearest neighbors. Below the transition to the XY ferromagnet, for ${{J_2^z < 1}}$, ${{\mathcal{S}_{ij}^z \approx -0.125}}$, indicating that the ground state is made up of a superposition of states in the 3:1 manifold, whilst for ${{J_2^z > 1}}$, ${{\mathcal{S}_{ij}^z \approx 0}}$, consistent with a superposition of states from the 2:2 manifold.
	}
	\label{fig:ed_j2_jp}
\end{figure*}

\section{Exact Diagonalization}
\label{sec:ED}

We now study the ground state of the XXZ Hamiltonian with ferromagnetic $J_{1,2}^{\perp}$ using the Lanczos method in exact diagonalisation (ED). 
In the following we set $J_1^z =1$ and tune the ratio $\gamma$ through $J_2^z=1/\gamma$. We shall present results for both $N=24$ and $N=36$ with periodic boundary conditions, with the former corresponding to the minimal cubic unit cell as shown in Fig.~\ref{fig:ed_24}. Fortunately, even though loops of length 4 are possible in these clusters, the corresponding perturbation Hamiltonian preserves the local \uo gauge transformation of Eq.~(\ref{eq:u1gauge}) (see App.~\ref{app:ED}). Therefore, although the low-energy physics will differ in the details due to finite size effects, our ED analysis shall give useful insights into the model, allowing us to go beyond perturbation theory, as well as to access the ground state for $\gamma \neq 2$.

In our calculations we use the fact that the Hamiltonian commutes with $m_z=\sum_i S_i^z$, diagonalizing the Hamiltonian in each $m_z$ sector separately, as well as lattice symmetries for the $N = 36$ cluster.
In the ground state, we measure the correlators
\begin{eqnarray}
	\mathcal{S}^z_{ij} = \langle S_i^z S_j^z \rangle, \qquad 
	\mathcal{S}^{\pm}_{ij} = \langle S_i^+ S_j^- + \mrm{h.c} \rangle,
\end{eqnarray}
up to 5th nearest neighbors.
The neighbors are illustrated in Fig.~\ref{fig:ed_24}.

To locate ground state phase boundaries, we use the second derivative of the ground state energy, which, for example, was used in Ref.~[\onlinecite{hickey2020}] to locate the boundaries of a gapless quantum spin liquid on small clusters.
At $T = 0$ the free energy reduces to the ground state energy, which is a function of the parameters of the Hamiltonian, $E_0 = E_0(\mbf{J})$.
Therefore non-analyticities in the ground state energy with respect to the parameters, which are detected via discontinuities in its derivatives, indicate a quantum phase transition~\cite{sachdev2011}.
As such, we measure the second derivative $\partial^2 E_0(\mbf{J}))/(\partial J_i^{\alpha})^2$,
at various points and along various directions of the parameter space $\mbf{J} = (J_1^z,J_1^{\perp},J_2^z,J_2^{\perp}$).
In particular, since the ground state energy is obtained on a discrete grid, we compute the finite central difference
\begin{eqnarray}
	c_i^{\alpha} (\mbf{J}) &=& \frac{\partial^2 E_0(\mbf{J})}{(\partial J_i^{\alpha})^2}
    \label{eq:second_derivative}\\
	&\approx& \frac{E_0(\mbf{J}+\delta\hat{\mbf{J}}_i^{\alpha}) - 2E_0(\mbf{J})+E_0(\mbf{J}-\delta\hat{\mbf{J}}_i^{\alpha})}{\delta^2},\nonumber
\end{eqnarray}
to approximate the second derivative.

We also find it useful to compute fidelities,
\begin{eqnarray}
    F = \abs{\braket{\Psi|\Phi}}^2,
    \label{eq:fidelity}
\end{eqnarray}
between normalized states.
For example we can compute the ground state fidelity relative to a reference point, $\mathbf{J}^*$, in the parameter space,
\begin{equation}
    F(\mathbf{J},\mathbf{J}^*) = \abs{\braket{\Psi_0(\mathbf{J})|\Psi_0(\mathbf{J}^*)}}^2,
    \label{eq:parameter_fidelity}
\end{equation}
as a measure of the change in the ground state wavefunction.
\subsection{Ground state phase diagram vs $J_2^z$}

First, we consider  $J_1^{\perp} = J_2^{\perp}=-J^\perp/\sqrt{2}$ and diagonalize the Hamiltonian on a $(J_2^z,J^{\perp})$ phase diagram displayed in Fig.~\ref{fig:ed_j2_jp}. In this convention, positive $J^\perp$ corresponds to the unfrustrated transverse ferromagnetic exchange.

For small $J_2^z$, up to $J_2^z\lesssim 1/3$, the ground state bears a finite magnetization as it is dominated by the proximity to the classical Ising ferrimagnet. This phase is rather trivial and will not be discussed further here. For $J_2^z \gtrsim 1/3$ we find that the ground state is in the $m_z=0$ sector, and the second derivative of the energy underlines that it is divided into three main phases (left panel of Fig.~\ref{fig:ed_j2_jp}).

At ``large'' $J^{\perp}$ the ground state is naturally the XY ferromagnet, as confirmed by the finite value of the $S_{ij}^\perp$ correlators in the middle panel of Fig.~\ref{fig:ed_j2_jp}. 
The XY ferromagnet noticeably persists down to $ J^{\perp}\rightarrow 0^+$ at $J_2^z=1$. This dip in the phase diagram can be rationalized as a consequence of the gigantic degeneracy enhancement at the classical boundary point $ J^{\perp}= 0$ and $J_2^z=1$.
There we know that all 3 up/1 down, 3 down/1 up and 2 up/2 down configurations are ground states \cite{nutakki2023}. 
It means that for each tetrahedron, 14 out of 16 states minimize the classical energy, making this boundary point one of the most degenerate classical ground states known in the literature, with a residual entropy of approximatively $\ln(5)/3\approx 0.536$  [\onlinecite{nutakki2023},\onlinecite{pohle2023a}].
Spin-spin correlations are thus very short ranged, and this classical boundary point is a quasi-paramagnet. 
In that sense, it is understandable that, within precision of the numerics, an infinitesimal transverse coupling $ J^{\perp}$ would suffice to polarize the system. For that reason, the XY ferromagnet separates the two other phases of Fig.~\ref{fig:ed_j2_jp} from each other for any finite value of $ J^{\perp}$.

For $J_2^z > 1$, the correlator $\mathcal{S}^{z}_{ij}$ between nearest neighbors is null (see the right panel of Fig.~\ref{fig:ed_j2_jp}), as expected from a 2:2 manifold. The 2:2 manifold persists up to a finite value of $J^\perp$ where the transverse correlator $\mathcal{S}^{\pm}_{ij}$ shows no sign of long-range order. We shall label this phase (2:2)$^{\perp}$, which we expect to be similar to the traditional quantum spin ice on pyrochlore  \cite{hermele2004,banerjee2008,shannon2012,huang2018} where center spins play little role.\\

Finally, we shall focus the rest of the discussion on the (3:1)$^{\perp}$ regime for $1/3\lesssim J_2^z < 1$. Although this state is fragile, surviving at its greatest extent up to about $J^{\perp}_c/J_1^z \approx 0.04$, this extent is comparable to the one of the \uo QSL on the pyrochlore, which exists up to $-J_{2,c}^{\perp}/J_2^z \approx 0.05$~\cite{huang2018}. 

\begin{figure}
\centering\includegraphics[width=0.8\linewidth]{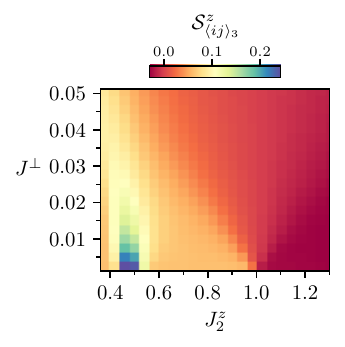}
\caption{
$\mathcal{S}_{ij}^z$ correlator between third neighbors (i.e. between center spins of two neighboring tetrahedra) for ${{J_1^z = 1, J_1^{\perp} = J_2^{\perp} = -J^{\perp}/\sqrt{2}}}$ in the ${{m_z = 0}}$ sector. While the correlator is finite for $J_2^z<1$, it shows a clear maximum at $J_2^z\sim 0.5$ for very small $J^\perp$. This indicates a possible ferromagnetic ordering of the center spins in the 3:1 regime.
}
\label{fig:ED24Sz3rdNN}
\end{figure}

\begin{figure*}[ht]
\centering\includegraphics[width=0.9\linewidth]{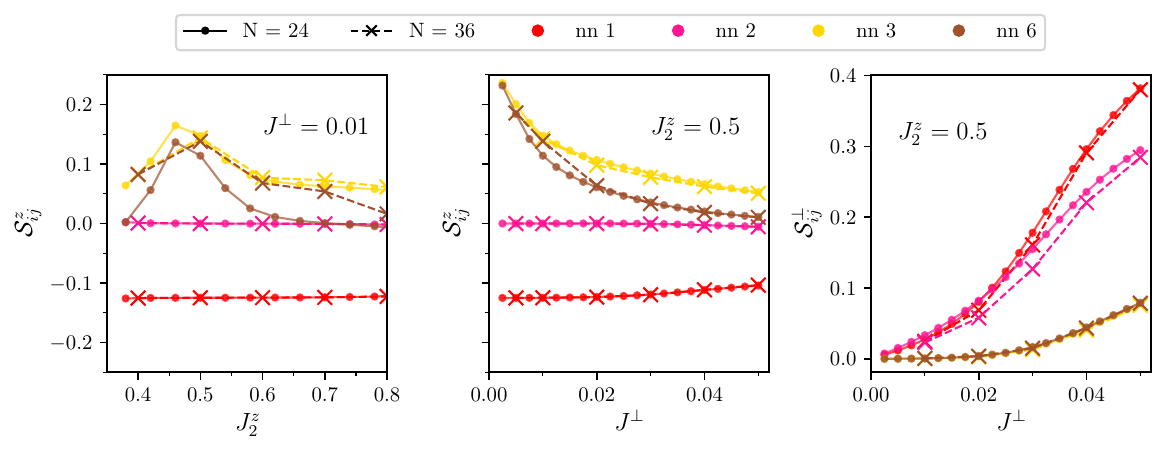}
\caption{
Correlators for ${{J_1^z = 1, J_1^{\perp} = J_2^{\perp} = -J^{\perp}/\sqrt{2}}}$ in the ${{m_z = 0}}$ sector comparing ED results for $N=24$ and $36$, showing results for some of the neighbors indicated in Fig.~\ref{fig:ed_24}. At $J_2^z=0.5$, the $\mathcal{S}_{ij}^{\pm}$ correlator (right) shows the decrease of XY ferromagnetic order as $J^\perp$ decreases, in favor of the onset of the ferromagnetic $\mathcal{S}_{\langle ij\rangle_3}^{z}$ correlators between nearest-neighbor center spins (middle, nn 3 and nn 6). In parallel, the $\mathcal{S}_{ij}^{z}$ correlator is consistent with 3:1 states where each center spin is surrounded by 1 parallel spin and 3 spins in the opposite direction; $\mathcal{S}_{\langle ij\rangle_1}^{z}$ saturates to -0.125 between neighboring center and vertex spins and $\mathcal{S}_{\langle ij\rangle_2}^{z}$ is zero between neighboring vertex spins.
(Left) At $J^\perp=0.01$, $\mathcal{S}_{ij}^z$ shows how the local (3:1)-tetrahedral correlations and ferromagnetic correlations between center spins persist over the entire regime $1/3\lesssim J_2^z \lesssim 0.8$. 
}
\label{fig:ED2436}
\end{figure*}

Based on Fig.~\ref{fig:ed_j2_jp}, the (3:1)$^{\perp}$ regime appears to be uniform. The transverse correlator $\mathcal{S}^{\pm}_{ij}$ is essentially zero and the longitudinal $\mathcal{S}_{ij}^z$ correlator between nearest neighbors becomes negative. Specifically, ${{\mathcal{S}_{ij}^z \approx -0.125}}$ between the center and vertex spins of a tetrahedron, as expected from a superposition of 3:1 states. However, in Fig.~\ref{fig:ED24Sz3rdNN}, while the $\mathcal{S}_{ij}^z$ correlator between two neighboring center spins is finite and positive for the entire region $1/3<J_2^z<1$, it shows a clear maximum around $J_2^z \approx 0.5 \Leftrightarrow \gamma\approx 2$, that persist upon increasing $J^\perp$.  ED results for $N=36$ essentially confirm the $N=24$ data [Fig.~\ref{fig:ED2436}]. This ferromagnetic correlation between center spins persist even up to 6$^{\rm th}$ nearest neighbor, i.e. second nearest neighbors on the diamond lattice (see Fig.~\ref{fig:ed_24} for nearest-neighbor definition and Fig.~\ref{fig:ED2436} for results).

\begin{figure*}
\centering
\begin{minipage}{6.5cm}
\includegraphics[width=\textwidth]{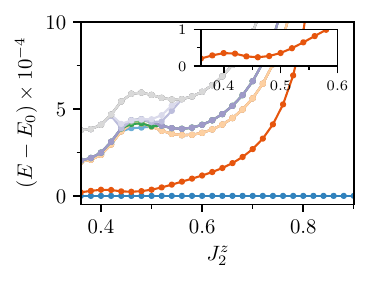}
\end{minipage}
\begin{minipage}{8cm}
\includegraphics[width=\textwidth]{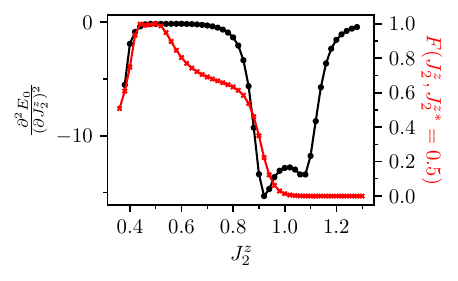}
\end{minipage}
\begin{textblock}{1}(1,-2.8)
	{(a)}
\end{textblock}
\begin{textblock}{1}(6,-2.8)
    {(b)}
\end{textblock}
\caption{
Low energy properties for ${{J^{\perp}=0.01}}$ as a function of $J_2^z$. \textbf{(a)} The 20 lowest energy eigenvalues relative to the ground state energy, with the gap to the first excited state shown in the inset. The gap has a local minimum at ${{J_2^z = 0.45}}$, where there is a cusp in the energies of the excited states. \textbf{(b)} The second derivative of the ground state energy and fidelity $F(J_2^z,J_2^{z*})$, as defined in eq.~\ref{eq:parameter_fidelity}. The second derivative has a clear double trough resulting from transitions to and from the XY ferromagnet, where the fidelity drops to zero.
At low $J_2^z$ the second derivative and fidelity decreases due to the nearby transition to the Ising ferrimagnet.
Within the region $0.4 \lesssim J_2^z \lesssim 0.9$ there are no clear signals in the second derivative, although there is some decrease in the fidelity away from $J_2^z = 0.5$.
}
\label{fig:ed_j2_jp=0.01}
\end{figure*}

A close look around $J_2^z=1/2$ along $J^{\perp} =0.01$ in Fig.~\ref{fig:ed_j2_jp=0.01} also brings to light subtle changes in the energy spectrum and ground state fidelity. 
The modification of the energy spectrum is to be expected because of the energy crossing of excited states and their asymmetry on either side of $J_2^z=1/2$ is consistent with the non-equivalence of the excited states: the two 4:0 states on one side and the twelve 2:2 states on the other side. 
The energy gap has a broad minimum around $J_2^z = 0.45$, which could be indicative of a quantum phase transition.
Furthermore, the fidelity decreases substantially away from $J_2^z = 0.5$ but has no sharp features.
On the other hand, the second derivative of the ground-state energy remains zero at the scale of Fig.~\ref{fig:ed_j2_jp=0.01}(b) (see black curve at $J_2^z\sim1/2$).
Overall, our results indicate short to long range ferromagnetic correlations between center spins in the entire (3:1)$^{\perp}$ regime for  $1/3<J_2^z<1$, which become noticeably stronger at $J_2^z\approx 0.5$ when the first excited states of a tetrahedron become degenerate (see Table \ref{tab:1}); this increase of correlations could be accompanied by a quantum phase transition around $J_2^z\approx 0.5$.\\

\subsection{Input from gauge theory} 
\label{sec:GT}

The 3:1 manifold comes from the $G_r = 0$ gauge sector [Eq.~(\ref{eq:Gdef})] and from the previous section, we know that the $G_r = 0$ constraint can be rewritten in gauge theory as $\mrm{div}(e)_r = q_r$ [Eq.~(\ref{eq:diveq})], where center spins play the roles of static charges. We can use the gauge theory 
to rationalize the ED results.

\begin{figure}[ht]
\centering\includegraphics[width=0.6\linewidth]{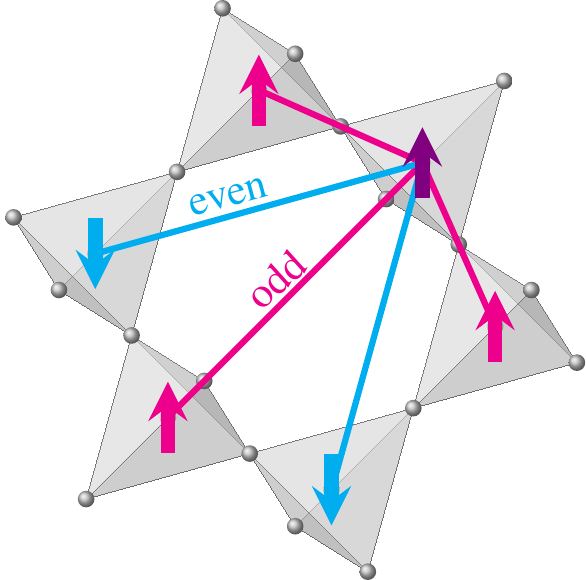}
\caption{
There are different ways for the virtual process $B_{\hexagon}$ of Fig.~\ref{fig:pert_4H3} to couple two center spins (gauge charges) across an hexagon. For any given center spin (violet), there are three spins at odd distance (magenta) and two spins at even distance (cyan); the $B_{\hexagon}$ would make the former parallel and the latter anti-parallel.
}
\label{fig:oddvseven}
\end{figure}

Assuming that  the gauge theory is deconfined in the $G_r = 0$ sector and Gaussian quantum electrodynamics is applicable, static charges will then interact via an effective Coulomb potential, at least at short distance. 
The charge background with the lowest classical energy would be the zinc-blende structure, with positive and negative charges alternatively positioned on the bipartite diamond lattice.
Because of the alternating value of $\eta_r=\pm 1$ in the definition of the charge in Eq.~(\ref{eq:diveq}), this zinc-blende charge structure corresponds to a ferromagnetic alignment of the central spins, which is consistent with our ED results [Fig.~\ref{fig:ED24Sz3rdNN}].
Furthermore, at third order in perturbation theory, $H_{p}^{(3)}$ does not couple sectors with different center spin configurations. 
Together, this implies a spontaneous breaking of spin inversion symmetry, choosing one of $S^z_i = + 1/2$ or $-1/2$ for all center spins.

At fourth order, matter acquires dynamics thanks to the $B_{\hexagon}^{(\dagger)}$ hopping term.
Fig.~\ref{fig:oddvseven} illustrates how the matter hopping behaves differently depending on the odd versus even distance between charges. 
Around an hexagonal loop, there are three (resp. two) possibilities for a pair of fermions to be at odd (resp. even) distance from each other; a straightforward counting argument thus favors the former. Furthermore center spins at even distance from each other form a network of spins sitting on a face-centered-cubic (fcc) lattice. Since the $B_{\hexagon}^{(\dagger)}$ hopping term requires antiferromagnetic orientations for center spins at even distance, the frustration on the fcc lattice strongly hinders charge hopping at even distance.
The $B_{\hexagon}^{(\dagger)}$ terms thus favor spin states allowing for odd-distance hopping, which corresponds to a ferromagnetic alignment of center spins [Fig.~\ref{fig:pert_4H3}(a) and \ref{fig:oddvseven}]. This offers an explanation for the ferromagnetic order we observe in ED that goes beyond the traditional third-order-perturbation picture inherited from the pyrochlore model.

Our emergent gauge field theory is thus consistent with ED, where all tetrahedra are in a 3:1 state and center spins are ferromagnetically correlated. At this stage, it is not possible, however, to know whether this correlation is short range or long range. The latter would then indicate a $\mathbb{Z}_2$ broken symmetry of the zinc-blende gauge-charge crystallization, consistent with magnetic fragmentation, where a \uo quantum spin liquid would co-exist with long-range order \cite{brooks-bartlett2014a}. All center spins would e.g. point up; the minority vertex spin of all tetrahedra would then be pointing up, and the three majority vertex spins would be pointing down. This phase would be similar to the magnetization plateau of spinels \cite{bergman2006a,bergman2006} mentioned previously.

\subsection{Input from quantum dimer models} 
\label{sec:QDM}

A complementary view to our spin model is as a variant of a quantum dimer model. Mapping the vertex spins to dimers, such that $S_i^z = +(-) 1/2$ corresponds to the presence(absence) of a dimer on the corresponding link of the diamond lattice. In order to respect the 3:1 constraint, if the center spin on a diamond site is $+1/2$, then there must be only one dimer touching that site, whereas if the center spin is $-1/2$, then the constraint is that three dimers touch that site [Fig.~\ref{fig:31regime}]. Therefore, at \textit{third} order in perturbation theory, Hamiltonian (\ref{eq:Hp3hex}) can be rewritten in the dimer language,
\begin{eqnarray}
H_{p}^{(3)}(\hexagon) = K_1 \sum_{\hexagon} \bigg(\ket{\hexa}\bra{\hexb} + \ket{\hexb}\bra{\hexa}\bigg),
\label{eq:Hp3dimer}
\end{eqnarray}
which acts on different Hilbert spaces depending on the central spin configuration, giving rise to what can be thought of as a family of separate dimer models. The Hilbert space can be further divided into sectors within which configurations are connected by $A_h^{(\dagger)} = \ket{\hexa}\bra{\hexb} \bigg(\ket{\hexb}\bra{\hexa}\bigg)$.

In terms of quantum dimers, if the center spins were ferromagnetically long-range ordered, the broken spin-inversion symmetry would result in having one and only one dimer on each tetrahedron (the minority vertex spin).
Our problem then comes down to the hard-core quantum dimer model on the diamond lattice. This model has been studied in Refs.~[\onlinecite{bergman2006,sikora2009,sikora2011}] where Hamiltonian (\ref{eq:Hp3dimer}) can be brought to the Rokhsar-Kivelson (RK) form~\cite{rokhsar1988} by adding a ``chemical potential",
\begin{eqnarray}
	H_{\mrm{RK}} &=& -\abs{K_1} \sum_{\hexagon} \bigg(\ket{\hexa}\bra{\hexb} + \ket{\hexb}\bra{\hexa}\bigg)\nonumber\\
	&+& \mu \sum_{\hexagon} \bigg(\ket{\hexa}\bra{\hexa} + \ket{\hexb}\bra{\hexb}\bigg).
	\label{eq:H_RK6}
\end{eqnarray}
At the RK point, $\abs{K_1} = \mu$, in each disconnected sector, the ground state is the equal superposition of all valid 3 up/1 down vertex spin configurations, a quantum dimer liquid. All of these ground states are degenerate at the RK point, but the degeneracy is broken away from it. Ref.~[\onlinecite{sikora2011}] numerically showed that the quantum liquid only survives down to $\mu_c = 0.75(2) \abs{K_1}$, where it transitions into an ordered state called the R state. This is adiabatically connected to the $\mu \rightarrow -\infty$ ground state of maximal number of flippable plaquettes. This suggests, though does not prove, that the ground state of $H_{p}^{(3)}$ is ordered.

This argument is, however, only valid at third order in perturbation theory. As discussed in section \ref{sec:GT}, fourth-order perturbation is consistent with ferromagnetic center spins, and thus, with the hard-core quantum dimer model. However, the $B_{\hexagon}^{(\dagger)}$ hopping term with charges at odd distance requires for vertex spins facing each other across an hexagon to point in the same direction [Fig.~\ref{fig:pert_4H3}(a)]. This is incompatible with R states where the maximization of flippable plaquettes favors opposite vertex spins to point in opposite directions \cite{sikora2011}. In other words, while the fourth-order virtual process $B_{\hexagon}^{(\dagger)}$ orders the center spins ferromagnetically -- at least on a finite length scale -- and thus stabilizes a hard-core quantum dimer model on the diamond lattice, it also induces a quantum dynamics that would melt the magnetic order suggested at third order. 
Hence, the R states do not seem to be a valid option for the ground state observed in our ED computations. However thanks to this connection to quantum dimer models, we can go one step further and build a quantum spin liquid variational wavefunction.

\begin{figure}
	\centering\includegraphics[width=0.9\linewidth]{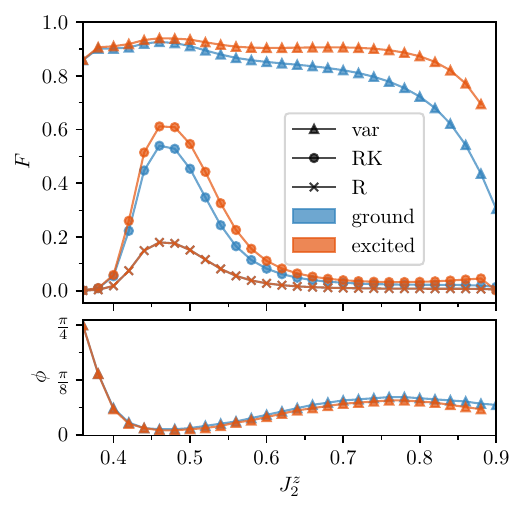}
	\caption{
	(Upper panel) Fidelity of the QSL variational wavefunction with the ground state (for ${{X = +1}}$) and first excited state (for ${{X = -1}}$) for $J_1^{\perp} = J_2^{\perp}=-0.01/\sqrt{2}$. For the data represented by triangles, $\phi$ is optimized to maximize the overlap, whereas for circles ${{\phi = 0}}$, i.e it is the pure RK-like wavefunction. The corresponding $\phi$ of the optimized function is shown in the lower panel.
    Fidelity with the ordered R state is also shown (crosses).
    \label{fig:fidRK}
	}
\end{figure}

\subsection{Rokhsar-Kivelson variational wavefunction}

To get more insight into the nature of the ground state in the (3:1)$^{\perp}$ region, we define a trial quantum spin liquid wavefunction which corresponds to the deconfined phase of the lattice gauge theory derived in perturbation theory. 
Our starting point is the equal weight superposition of all 3:1 configurations compatible with all center spins pointing up, $\ket{RK(\uparrow)}$. Note that the third order length-6 ring-exchange term on the infinite lattice does not connect all of these configurations, however the second-order length-4 ring-exchange term (which exists on the finite size cluster) does.
Now, we also want to include the effect of the $J_1^{\perp}$ term which enters at third-order on length-4 loops and fourth-order on length-6 loops.
We observe that $\sum_{\hexagon} (B_{\hexagon}+B_{\hexagon}^{\dagger})\ket{RK(\uparrow)}$, where $\hexagon$ represents all length-6 loops, both winding and non-winding, introduces all of the configurations included by $(B_{\square}+B_{\square}^{\dagger})$ (due to periodic boundary conditions on the finite-size cluster) plus some additional ones.
We can thus restrict our ansatz to $B_{\hexagon}^{(\dagger)}$ dynamics only, without having to add artificial terms due to finite-size effects.
Furthermore, since the full Hamiltonian commutes with the spin inversion operator, $X$, we want to ensure that the wavefunction is spin-inversion symmetric ($X=+1$) or antisymmetric ($X=-1$). Therefore, our variational wavefunction is
\begin{eqnarray}
\Psi_v(X,\phi) \propto \bigg(\cos \phi + \sin \phi \sum_{\hexagon} \bigg(B_{\hexagon}+B_{\hexagon}^{\dagger}\bigg) \bigg) \nonumber\\
\times (\ket{RK(\uparrow)} \pm \ket{RK(\downarrow)}),
\label{eq:RKvar}
\end{eqnarray}
where $\phi$ is the variational parameter which may be optimized to maximize overlap with the ground state.

Since with $J_1^{\perp},J_2^{\perp}<0$, the Hamiltonian is stoquastic, the ground state cannot have $X = -1$. But we chose to introduce this as we find that the $X = -1$ state has high overlap with the first excited state.

Armed with this ansatz, and fixing $J^{\perp} = 0.01$ which allows us to explore the $(3:1)^\perp$ regime over a broad range of parameters [Fig.~\ref{fig:ed_j2_jp}], we compute the fidelity as a function of $J_2^z$ in Fig.~\ref{fig:fidRK}. The ground state fidelity with the optimised variational wavefunction is over $80\%$ for a substantial portion of the parameter space around $J_2^z = 0.5$, decreasing as $J_2^z$ approaches the phase transition to the XY ferromagnet near $J_2^z=1$. The optimal $\phi$ is minimum around $J_2^z = 0.47$, which coincides with the local minimum in the energy gap of Fig.~\ref{fig:ed_j2_jp=0.01}(a) and the sharp increase of ferromagnetic correlations between center spins in Fig.~\ref{fig:ED24Sz3rdNN}.

\begin{figure*}
	\centering
	\includegraphics[width=16cm]{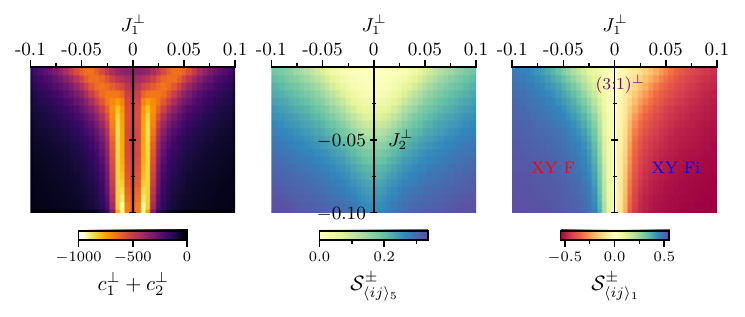}
	\caption{ED ground state properties for ${{J_1^z = 1, J_2^z = 0.5}}$ with positive and negative $J_1^\perp$ ($x-$axis) and negative $J_2^\perp$ ($y-$axis), for the $N = 24$ cluster and $m_z = 0$ sector. (Left) The sum of the second derivatives, ${{c_1^{\perp} + c_2^{\perp}}}$ computed using finite differences [Eq.~(\ref{eq:second_derivative})]. (Middle) Correlator $\mathcal{S}_{ij}^{\pm}$ between 5$^{\rm th}$ nearest neighbors (both are vertex spins), showing the onset of long-range ferromagnetic XY correlations between vertex spins for larger $J^{\perp}$. (Right) Correlator $\mathcal{S}_{ij}^{\pm}$ for nearest neighbors (center and vertex spins). This shows how for antiferromagnetic $J_1^{\perp}$ the ordered state at larger $J^{\perp}$ is an XY ferrimagnet, with centers and vertices anticorrelated. This must be the case given the transformation properties of the Hamiltonian discussed in section \ref{sec:Hamsym}. For small $J_1^{\perp}$ there is an intermediate region where the $S^\pm$ components of center and vertex spins are uncorrelated.}
	\label{fig:ed_j1_j2}
\end{figure*}

Hence, the ground state and first excited state are quasi-degenerate at $J_2^z\approx 0.5$ and well represented by a simple $\ket{RK(\uparrow)} \pm \ket{RK(\downarrow)}$ wavefunction. This suggests a possible quantum critical point, with the caveat that the ground state wavefunction is liquid-like on either side of this point.

We also evaluate fidelities for the ordered R state (see eq.~\ref{eq:R_fidelity}), in order to rule out the possibility of the system adopting this competing ordered state. These remain small ($< 20\%$) in this region of the parameter space, indicating that the low energy states contain additional fluctuations of the center and vertex spins.
Note that for this $N = 24$ cluster, the fidelity of the R state with the $\phi=0$ variational wavefunction is $1/3$, since $1/3$ of the basis states making up the superposition in $\ket{RK(\uparrow)}$ are also included in the R state (and the same for $\ket{RK(\downarrow)}$, see Eq.~\ref{eq:R_fidelity}).

\subsection{Ground state phase diagram for $J_1^\perp \neq J_2^\perp$}

\begin{figure}
	\centering \hspace{-0.5cm} \includegraphics[width=6.5cm]{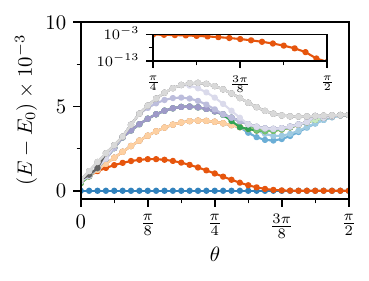}
	\centering\includegraphics[width=8cm]{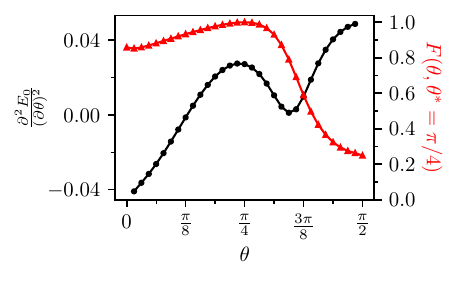}
	\caption{Low energy properties for ${{J_2^z = 0.5}}$, ${{J^{\perp}=0.02}}$, whilst varying $\theta$. (Upper) The 20 lowest energy eigenvalues relative to the ground state energy, with the gap to the first excited state shown in the inset on a log scale. The gap becomes small (of order $10^{-5}$) at ${{\theta \approx 3\pi/8}}$ where there is also a local minimum in the energies of the excited states. (Lower) The second derivative of the ground state energy and fidelity $F(\theta,\theta^*)$, as defined in Eq.~(\ref{eq:parameter_fidelity}). The second derivative shows a small dip near $\theta = 3\pi/8$, around which the fidelity decreases significantly.}
	\label{fig:ed_j2=0.5_jp=0.02}
\end{figure}

To conclude, let us study the distinct influences of $J_1^\perp$ and $J_2^\perp$ separately. We fix $J_2^z = 1/2$ and parametrize the perturbation couplings as
\begin{eqnarray}
\begin{split}
J_1^{\perp} &= -J^{\perp} \cos \theta,\\
J_2^{\perp} &= -J^{\perp} \sin \theta.
\end{split}
\end{eqnarray}
In Fig.~\ref{fig:ed_j1_j2}, we find the anticipated XY ferromagnet ground state for ferromagnetic $J_1^{\perp}$, which becomes a ferrimagnet for antiferromagnetic $J_1^{\perp}$, as a result of the transformation properties of the Hamiltonian discussed in section \ref{sec:Hamsym}.
The (3:1)$^{\perp}$ state is slightly more stable to $J_1^{\perp}$ than $J_2^{\perp}$ terms, likely a result of the larger coordination number of $J_2^{\perp}$ (6 vs 4). In Fig.~\ref{fig:ed_j2=0.5_jp=0.02}, we look at a cut along $J^{\perp} = 0.02$, varying $\theta$ between $0$ and $\pi/2$. 
For $\theta$ small, $J_2^{\perp} \approx 0$, so non-constant terms appear only at sixth order (fourth order on the small clusters) in perturbation theory where the ring-exchange is entirely mediated by $J_1^{\perp}$. At this order, another type of ring exchange exists which flips all center spins around the hexagonal (square) loop. 
It means that spinon dynamics is allowed, but the high order in perturbation theory makes the energy gaps comparatively small as $\theta\rightarrow 0$ in Fig.~\ref{fig:ed_j1_j2}. As we increase $\theta$, we notice the fidelity relative to $\theta^* = \pi/2$ decreases and the second derivative of the energy has a dip around $\theta \sim 3\pi/8$ accompanied by a quasi-degeneracy of the first excited state, whose gap becomes smaller than numerical precision at $\theta = \pi/2$ (where $J_1^{\perp} = 0$). 
This is consistent with the fact that, as $J_1^{\perp} \rightarrow 0$, quantum fluctuations of center spins vanish and only loops of vertex spins are possible.
The Hilbert space splits into disconnected blocks accompanied by a quasi-degeneracy.

\begin{figure}
	\centering\includegraphics[width=0.8\linewidth]{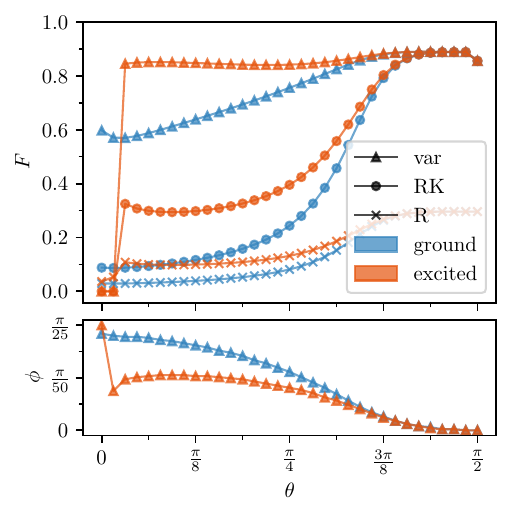}
	\caption{
	(Upper panel) Fidelity of the QSL variational wavefunction (var) with the ground state (for ${{X = +1}}$) and first excited state (for ${{X = -1}}$) for ${{J_2^z =0.5, J^{\perp} =0.02}}$. For the data represented by triangles $\phi$ is optimized to maximize the overlap, whereas for circles ${{\phi = 0}}$, i.e it is the pure RK-like wavefunction. The corresponding $\phi$ of the optimized wavefunction is shown in the lower panel.
    Fidelity with the ordered R state is also shown.
	}
    \label{fig:fidRKtheta}
\end{figure}

Something non-trivial is happening at $\theta \sim 3\pi/8$. We take advantage of the variational wavefunction of Eq.~(\ref{eq:RKvar}) to explore this phenomenon, and compute the overlaps of the ground and first-excited states for $J_2^z = 0.5$ and $J^{\perp} = 0.02$, allowing $\theta$ to vary. Since $J_2^z = 0.5$, the perturbation theory is directly applicable here. We find that for $\theta \gtrsim 3\pi/8$ the overlap with the variational wavefunction is particularly high (at around 90$\%$) for $\phi=0$. This is consistent with perturbation theory since a very small $J_1^{\perp}$ means that correlations induced by higher-order terms than the $J_2^{\perp}$ ring-exchange (such as $B_{\hexagon}$) are minimal.
The closing of the energy gap, and high overlap with the RK wavefunction, point towards a ground state well-described by a deconfined phase dominated by $J_2^{\perp}$ terms in perturbation theory, as on the pyrochlore lattice. 
However, as discussed in section \ref{sec:QDM}, the ground state in the corresponding effective dimer model actually orders in the thermodynamic limit \cite{sikora2011}. This would suggest that the absence of order for small $J_1^\perp$ in ED might be a finite size effect.

On the other side of the phase diagram, as $\theta$ is decreased, i.e turning on $J_1^{\perp}$, the optimal $\phi$ increases, as the role of the $B_{\square,\hexagon}$ becomes more significant [Fig.~\ref{fig:fidRKtheta}].
The point where the optimal $\phi$ increases almost coincides with the opening of the energy gap and anomalies in the second derivative of the energy and fidelity, so these features may be associated with a change in the ground-state wavefunction to include $B_{\hexagon}$-induced correlations.
The overlap with ground state drops to below $60\%$ at small $\theta$, which is likely the result of only including $(B_{\hexagon}+B_{\hexagon}^{\dagger})$ at first order in the variational wavefunction, since higher-order $J_1^{\perp}$ terms would become significant.

Fidelity with the ordered R state for all values of $\theta$ remains relatively small, increasing to 1/3 as $\theta \rightarrow \pi/2$, consistent with the ground state wavefunction becoming well approximated by the pure RK ($\phi=0$) variational wavefunction.
The fact that the R state is not the ground state as $J_1^{\perp} \rightarrow 0$ could be a result of the length-4 ring-exchange terms allowed on the finite size cluster which are ergodic within the 3:1 and 1:3 sectors and so favor the inclusion of spin configurations outside of those contained within the R state.

Note that, whether one varies $\theta$ or $J_2^z$, in both instances the closing of the energy gap coincides with a larger overlap with the RK wavefunction and with noticeable features the second derivative of the energy, suggesting possible quantum critical points.

Using exact diagonalization, we have located the region in parameter space where the ground state consists of a 3:1 manifold with quantum fluctuations. The high ground-state overlap with a QSL variational wavefunction motivated from perturbation theory, even away from $J_2^z = 0.5$, suggests that the effective lattice gauge theory and quantum dimer models derived in perturbation theory offer the appropriate perspective from which to understand the low-energy properties. 
Due to the small system size, we are not able to establish whether this regime is a single phase or hosts several different phases. However, we identify points at $J_2^z \approx 0.5$, $J^{\perp} \approx 0.01$, $\theta \approx \pi/4$, and $J_2^z \approx 0.5$, $J^{\perp} \approx 0.02$, $\theta \approx 3\pi/8$ as candidates for forming part of quantum critical boundaries.
Numerical simulations on larger system sizes are needed to resolve the nature of these points, as well as to determine whether a liquid state survives to larger system sizes.
Crucially, a system size large enough so that the shortest closed loops on the lattice are the hexagons described in the perturbation theory would be necessary to properly probe the quantum spin liquid proposed by perturbation theory.


\section{Experimental signatures}
Let us consider what the experimental signatures of the QSL described in the previous sections could be, with the specific scenario in mind of going from a paramagnet, to its parent \zt classical spin liquid and finally into the quantum spin liquid as temperature is lowered.

Upon entering the \zt classical spin liquid, all tetrahedra are 3-up/1-down or 3-down/1-up (assuming no order of the centre spins) and there are zero correlations between vertex spins~\cite{slobinsky2018}.
Center spins would be ferromagnetically correlated at short distance, so an experimental signature of the classical phase would be an equal-time structure factor where 2/3 of the magnetic degrees of freedom are featureless, while 1/3 is short-range ferromagnetic. 
To differentiate this absence of correlations from traditional paramagnetism, one could check that it persists down to temperatures much smaller than the Curie-Weiss temperature estimated from magnetic susceptibility. 

Turning to the quantum phase, our ED results indicate that center spins should order, at least partially. 
If there is long-range ferromagnetic order of the center spins, then one would see Bragg peaks at the $\Gamma$ point. 
On the level of classical Heisenberg spins (i.e ignoring quantum effects), a ferromagnetic order of center spins, implies that vertex spins subject to the constraint $\gamma \mathbf{S}_{t,c} + \sum_{v=1}^4 \mathbf{S}_{t,v} = 0$ will form a Coulomb phase~\cite{nutakki2023} would give rise to pinch points in the structure factor, also at $\Gamma$ points. 
This is a typical example of magnetic fragmentation \cite{brooks-bartlett2014a}. 
Including quantum effects these pinch points should vanish in the elastic sector (as $\omega$ goes to zero in inelastic neutron scattering) with a linear (photon-like) dispersion relation~\cite{benton2012} (giving for example a $\sim T^3$ specific heat dependence at low temperature). 

On the other hand, if there is only short-range order of the center spins the Bragg peaks would broaden to indicate the finite ferromagnetic correlation length of the center spins.
The constraint to realize a \uo Coulomb phase would not be satisfied anymore, therefore pinch points would become diffuse and might be impossible to recognize.

\section{Summary and Outlook}
We studied the ground state properties of the quantum spin$-1/2$ XXZ model on the centered pyrochlore lattice, focussing on the region of ferromagnetic $J^{\perp}$.
In perturbation theory, for $\gamma = 2$, we derived an effective frustrated compact \uo lattice gauge theory coupled to fermionic matter. The deconfined phase of this theory, without condensation of spinon pairs, is a \uo quantum spin liquid.
The gauge theory differs from conventional compact lattice quantum electrodynamics by the coupling of matter and gauge fields, with a two-body interaction coupled by a loop of gauge fields.
Turning to exact diagonalization, we found a region in the parameter space, for small $J^{\perp}$, where the ground state is a superposition of 3:1 states, a prerequisite for realizing the QSL. We saw that the ground state in this regime has high overlap with a QSL variational wavefunction inspired by the Rokhsar-Kivelson point and perturbation theory, even away from the $\gamma = 2$ point where it was derived. This indicates that the perturbation theory and associated dimer models, provide a good framework for understanding the low-energy properties of the XXZ centered pyrochlore model. However, these computations suffer from finite size effects, and simulations on larger system sizes are called for to settle the precise nature of the phase(s) in the (3:1)$^{\perp}$ region.

Looking ahead, exact diagonalization on larger system size appears very challenging. However, this could possibly be improved if working in the reduced Hilbert space of the effective perturbative Hamiltonian, similar to~\cite{pace2021a}. Furthermore, such an approach could also be used to simulate the perturbative regime for $J^{\perp}$ antiferromagnetic, where, on the pyrochlore, the ground state is a $\pi$-flux \uo QSL. This, however comes at the cost of only being valid for $\gamma = 2$, and not for the broader parameter space.

Alternatively, one could use quantum Monte Carlo, which does not suffer from the sign problem for ferromagnetic $J^{\perp}$, to study the full XXZ Hamiltonian, similar to the calculations performed in ~\cite{huang2018}. To identify the QSL, one could, for example, compare correlation functions to those of the RK wavefunction for an appropriate configuration of center spins. One possible ordered state which would need to be ruled out, is the R state~\cite{sikora2011,bergman2006,bergman2006a} found at smaller $\mu$ in the effective third order dimer model for ferromagnetic centers. 
Whilst we found that this is not the ground state in the regions of the phase diagram we looked at using ED, this could be due to the presence of length-4 loops on the small clusters studied.
Indeed, we expect it to be the ground state for parameters $J_2^z = 0.5, J_1^{\perp} = 0, \abs{J_2^{\perp}} \ll 1$, and could occupy a finite extent in the parameter space.

If a QSL ground state is identified in the (3:1)$^{\perp}$ regime, it would then be interesting to characterize the properties of the state by interpreting the effective lattice gauge theory as an emergent quantum electrodynamics.
At third order in perturbation theory, the effective theory is quantum electrodynamics in a static charge background, but it is not clear how the interactions introduced at fourth order between dynamical spinons and gauge fields impacts this.
Within this picture, one could imagine, upon cooling, going from a high-temperature paramagnet to a classical \zt spin liquid of the Ising model via a crossover, followed by a phase transition into a \uo quantum spin liquid.

On the experimental side, a material whose low temperature properties are described by the XXZ model with ferromagnetic $J^{\perp}$ on the centered pyrochlore lattice is probably unrealistic, however, more promising, is the prospect of realizing a spin $1/2$ material with entirely antiferromagnetic interactions.
Since the perturbation theory is also valid for $J^{\perp}$ antiferromagnetic, there could also be a \uo quantum spin liquid occupying a finite extent in the space of antiferromagnetic interactions.
This could be possible, for example, in a Cu(II)-based metal-organic framework. 
More speculatively, perhaps it could be possible to find effective $S=1/2$ degrees of freedom in a variant with strong spin-orbit coupling, as in the dipolar-octupolar pyrochlores~\cite{huang2014}. 
Whether such a QSL exists, its spectroscopic and thermodynamic properties, as well as its extent in the parameter space, would be interesting questions to investigate.

This work is the first step towards understanding possible quantum spin liquids on the centered pyrochlore lattice and their effective gauge theory descriptions, which enriches the traditional \uo spin liquids on pyrochlores via their coupling to fermionic matter and subsequent quantum dynamics. 
\nocite{data}

\section*{Acknowledgments}
We wish to thank B. Sbierski for useful comments on the manuscript and J. Hall{\'e}n, M. Hermele, J. Knolle, and Y. Iqbal for insightful discussions.
R.P.N. acknowledges financial support from ANR-23-CE30-0018.
L.D.C.J.\ acknowledges financial support from ANR-23-CE30-0038-01 and thanks LMU for their hospitality in August 2024. L.P. acknowledges financial support from the Deutsche Forschungsgemeinschaft (DFG), project Nr 530111096.
S.C. thanks CALMIP (grant 2024-P0677) and GENCI (project A0150500225) for computer resources.
\section*{Data Availability}
The data that support the findings of this artice are openly available~\cite{data}.
%
%
\appendix	 	
\label{sec:appendix}
\section{Exact diagonalisation}
\label{app:ED}

ED calculations were performed on an $N = 24$ or $36$ site clusters with periodic boundary conditions. Due to periodic boundary conditions, the lattice contains closed loops of length 4, in addition to the usual length-6 loops on the infinite centered pyrochlore lattice.

\begin{figure*}
	\centering\includegraphics[width=15cm]{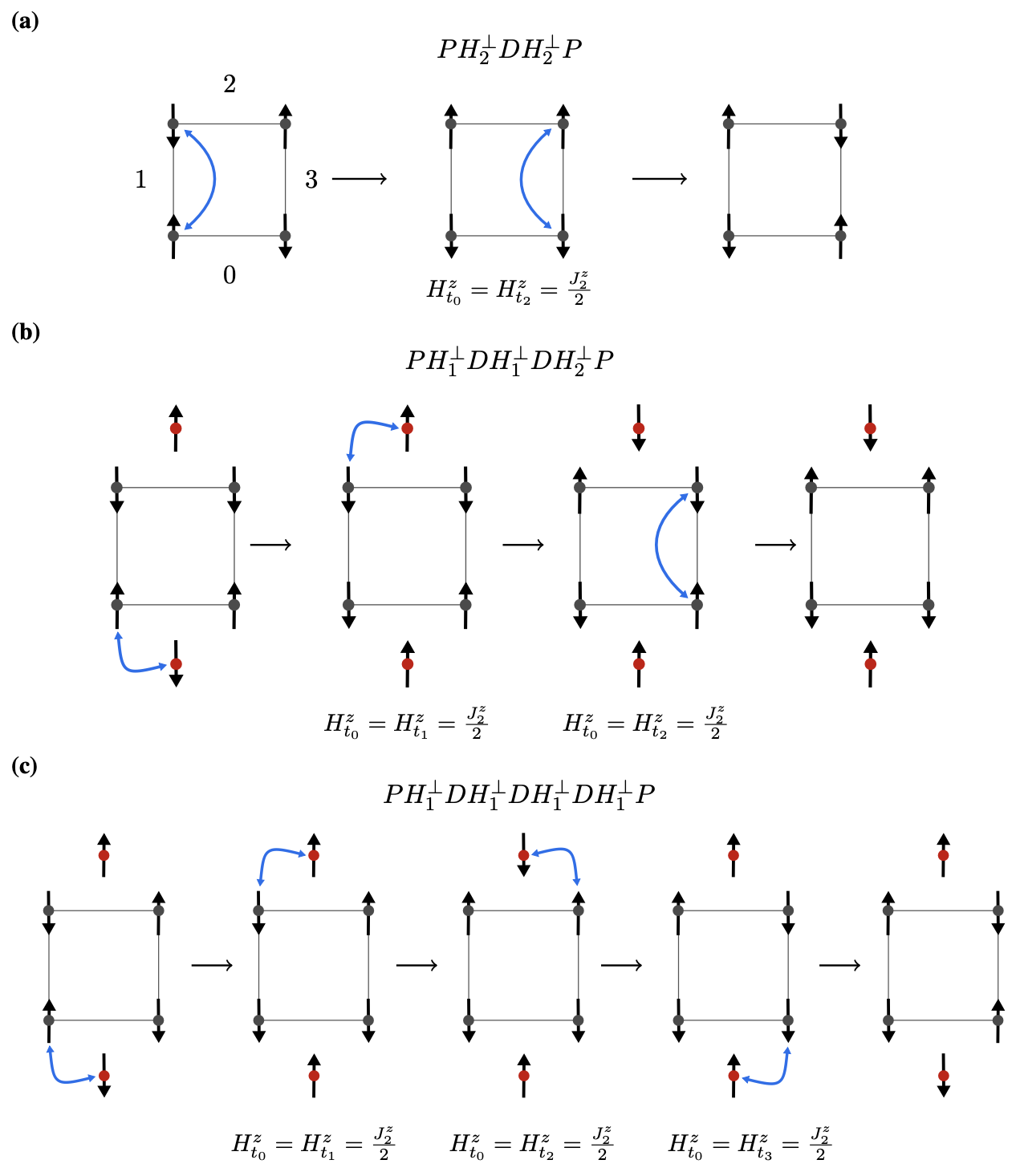}
	\caption{Loop processes in perturbation theory which arise in clusters containing length-4 closed loops. Note that the square shown is not a tetrahedron, the spins making up the loop belong to four tetrahedra around the outside of the loop (not explicitly shown see Fig.~\ref{fig:ed_24}.
	The labels, $i$, used to label the tetrahedra, $t_i$, are indicated in \textbf{(a)}.}
	\label{fig:ed_pert}
\end{figure*}
This modifies the $\gamma = 2$ perturbation theory of \ref{sec:pert_theory} by introducing non-constant lower order loop processes, which are illustrated in Fig.~\ref{fig:ed_pert}.
At second order, the non-constant contribution to $PH_2^{\perp}DH_2^{\perp}P$ is
\begin{eqnarray}
	H_{p}^{(2)}(\square) 
	&=& K^{\square}_1 \sum_{\square} \bigg(S_{v,0}^+ S_{v,1}^- S_{v,2}^+ S_{v,3}^- + \mrm{h.c}\bigg)\\
	&=& K^{\square}_1 \sum_{\square} \bigg(A_{\square}+ A_{\square}^{\dagger} \bigg),
	\label{eq:Hp2ed}
\end{eqnarray}
where $A_{\square} = S_{v,0}^+ S_{v,1}^- S_{v,2}^+ S_{v,3}^-$ and $K^{\square}_1 \propto -(J_2^{\perp})^2/J_2^z$, which is the analogue of the third order hexagonal ring exchange in Eq.~(\ref{eq:Hp3ring}).

At third order, one obtains the analogue of the ring-charge exchange involving center spins [Eq.~(\ref{eq:Hp4})],
\begin{eqnarray}
	\begin{gathered}
		H_{p}^{(3)}(\square) = K_2^{\square} \sum_{\square} \bigg( B_{\square} + B_{\square}^{\dagger} \bigg), \qquad
		B_{\square} = \sum_{i=0}^3\sum_{j=i+1}^3 B^{ij}_{\square},\\
		B^{ij}_{\square} = 
		\begin{cases}
			S_{c_i}^+ S_{c_j}^- S_{v_i}^- \dots S_{v_{j-1}}^+ S_{v_j}^+ \dots S_{v_{i+3}}^- \: \mrm{for} \: j-i \: \mrm{even}\\
			S_{c_i}^+ S_{c_j}^+ S_{v_i}^- \dots S_{v_{j-1}}^- S_{v_j}^- \dots S_{v_{i+3}}^- \: \mrm{for} \: j-i \: \mrm{odd}
		\end{cases},
	\end{gathered}
\end{eqnarray}
where $K_2^{\square} \propto J_2^{\perp}(J_1^{\perp})^2/(J_2^z)^2$.
In Fig.~\ref{fig:ed_pert}, we also show the fourth order non-constant $PH_1^{\perp}DH_1^{\perp}DH_1^{\perp}DH_1^{\perp}P$, which has the same effect as \ref{eq:Hp2ed}, with a pair of $J_1^{\perp}$ terms acting as a $J_2^{\perp}$ term.
Therefore, even with $J_2^{\perp} = 0$, one still has a non-zero ring-exchange type term in the effective Hamiltonian.
On the infinite centered pyrochlore lattice, this $J_1^{\perp}$ only ring-exchange first appears at sixth order.

Including $H_{p}^{(2)}(\square)$ and $H_{p}^{(3)}(\square)$ in the effective Hamiltonian does not change its symmetry properties, one still obtains a \uo lattice gauge theory after mapping to rotors and fermions.

In order to assess the properties of the ground (and first excited state) with respect to the competing ordered R state~\cite{bergman2006a,sikora2011}, we compute the sum of projectors
\begin{equation}
    F_R = \sum_{i=1}^{16} \abs{\braket{R_i|\Psi}}^2.
    \label{eq:R_fidelity}
\end{equation}
This can also be interpreted as the sum of fidelities for the 16 basis states corresponding to an R state.
These consist of 8 with all center spins up and corresponding 1:3 vertex spin configurations which have a flippable hexagonal plaquette in the cubic unit cell (the degenerate states in the dimer model of ~\cite{bergman2006a,sikora2011}), plus an additional 8 with all center spins down and corresponding 3:1 vertex spin configurations also with a flippable hexagonal plaquette.
\clearpage
\bibliography{Bibliography}
\end{document}